\documentclass[aps,reprint,pra]{revtex4-1}
\usepackage{graphicx}
\usepackage{dcolumn}
\usepackage{bm}
\usepackage{amsmath}
\usepackage{latexsym}
\usepackage{booktabs}
\numberwithin{equation}{section}

\begin{document}

\def\bfr{{\bf r}}
\def\bfk{{\bf k}}
\def\bfv{{\bf v}}
\def\conc{[c]}
\def\conu{[u]}
\def\calP{{\cal P}}
\def\calF{{\cal F}}
\def\calE{{\cal E}}
\def\calEavg{{\langle{\cal E}[c,\bfv]\rangle}}
\def\calS{{\cal S}}
\def\calT{{\cal T}}
\def\calL{{\cal L}}
\def\calD{{\cal D}}
\def\cgmodes{{\rm slow\  modes\ }}

\def\parsparev{{\Bigl({\partial S_r \over \partial E_r}\Bigr)}_V}
\def\parepartv{{\Bigl({\partial E_r \over \partial T}\Bigr)}_V}
\def\parepartvInline{{({\partial E_r/\partial T})}_V}
\def\pareparvt{{\Bigl({\partial E_r \over \partial V}\Bigr)}_T}
\def\partparpsr{{\Bigl({\partial T \over \partial P_r}\Bigr)}_{S_r}}
\def\partparpsrInline{{({\partial T/\partial P_r})}_{S_r}}
\def\dtdps{\Bigl ({\partial T\over \partial P}\Bigr )_S}
\def\dtdpsInline{({\partial T/\partial P})_S}
\def\parprpartv{{\Bigl({\partial  P_r\over \partial T}\Bigr)}_V}
\def\parprparvt{{\Bigl({\partial  P_r\over \partial V}\Bigr)}_T}
\def\parsparet{\Bigl ({\partial S\over \partial E_t}\Bigr)_{V,\conc}}
\def\parepsparps{({\partial\epsilon/\partial P})_S}

\def\ecgavg{\langle{\cal E}\conc\rangle}
\def\ecgvavg{\langle{\cal E}[c,\bfv]\rangle}
\def\part{\partial}
\def\decgdt{{\partial \ecgavg \over \partial t}}
\def\decgvdt{{\partial \ecgvavg \over \partial t}}
\def\alfp{\alpha_{P}}
\def\cp{C_P}
\def\cv{C_V}
\def\kapt{K_{T}}
\def\kaptr{K_{Tr}}
\def\kaps{K_{S}}
\def\kapsr{K_{Sr}}
\def\alfpr{\alpha_{Pr}}
\def\cpr{C_{Pr}}
\def\cvr{C_{Vr}}
\def\cpcg{C_{P{\rm sm}}}
\def\cvcg{C_{V{\rm sm}}}
\def\Ccg{C_{\rm sm}}
\def\Fcg{F_{\rm sm}}
\def\Pcg{P_{\rm sm}}
\def\dtdt{{dT\over dt}}
\def\dvdt{{dV\over dt}}
\def\dpdt{{dP\over dt}}
\def\dprdt{{dP_r\over dt}}
\def\depsdt{{d\epsilon\over dt}}
\def\shat{{\hat S}}
\def\lhat{{\hat L_\Lambda}}
\def\chit{\chi_T}
\def\uu2{\langle u^2\rangle}
\def\yy2{\langle y^2\rangle}
\def\stil{{\tilde S}}
\def\etil{{\tilde e}}
\def\alfstar{\alpha^\ast}
\def\dTcdp{{d T_c\over dP}}
\def\epsabs{{\vert\epsilon\vert}}
\def\epsfabs{{\vert\epsilon_f\vert}}
\def\epsoverepsf{{\epsilon/\vert\epsilon_f\vert}}

\title{Kinetics of phase separation in thermally isolated critical binary fluids}
\author{James P. Donley}
\affiliation{Valence4 Technologies, Eugene, OR 97405}
\email{jdonley@valence4.com}

\date{\today}

\begin{abstract}
Spinodal decomposition in a near-critical binary fluid is examined for experimental scenarios in which the liquid is quenched abruptly by changing the pressure
and the subsequent phase separation occurs with no heat flow from the outside,
i.e., adiabatically. Equations of motion for the system volume and effective temperature are derived. It is shown that for this case that the nonequilibrium decomposition process is well approximated
as one of constant entropy, i.e., as thermodynamically reversible.
Quantitative comparison, with no adjustable parameters, is made with 
experimental light scattering data of Bailey and Cannell [$\rm {Phys.\ Rev.\ Lett.\ }{\bf 70}$, 2110 (1993)].
It is found that including these adiabatic effects accounts for most of the discrepancies between these experiments and previous isothermal theory.
The equilibrium static critical
properties of the isothermal theory are also examined,
this discussion serving to justify some approximations in the current theory.
\end{abstract} 


\maketitle

\section{\label{sec:intro}Introduction}
Spinodal decomposition is the process of phase separation of a thermodynamically unstable mixture.\cite{cahn68,bray94,onuki02,desai09} It and its complement, nucleation, are two of the most common mechanisms of phase transformation for systems governed by a conserved order parameter.
Decomposition is a common way to create alloy materials, particularly metals, commercially, and has been predicted under conditions thought to occur in the early universe\cite{boyanovsky06}.

Decomposition was studied initially for systems in which phase separation
is driven by single particle diffusion, such as metal alloys.
The first theories of the early stage of decomposition in such substances
were due to Cahn and Hilliard,\cite{cahn68} and Cook\cite{cook70}. Later, Langer, Bar-on and Miller (LBM)\cite{langer75}, employed the more formal methods of the master equation vein of the theory of stochastic processes.\cite{vankampen81} More recent research has used the framework of quantum thermodynamics to explore this phenomenon.\cite{yamada19} Experimental tests of any of these theories have not been entirely unambiguous though. 
For metal alloys, lattice mismatch of
the two components can cause stresses to build during unmixing, which slows the
rate of decomposition. These strains can be minimized by matching the
lattice constants of the individual components \cite{mainville97}, or
 avoided entirely by examining unmixing in liquids \cite{chou79,wong78}.
In liquids though, unmixing is greatly accelerated by advection.
Kawasaki and Ohta (KO) extended the LBM theory to binary liquids
by incorporating these hydrodynamic effects.\cite{kawasaki78}

 A careful set of experiments to test this KO theory was done by
 Bailey and Cannell (BC) using 3-methylpentane and nitroethane (3MP+NE) in the
 critical region.\cite{bailey93,baileyThesis} The critical equilibrium properties of
 3MP+NE have been well characterized. Further, the components of this binary liquid have very similar
indices of refraction, which minimizes multiple scattering effects during
decomposition. Since all the parameters of the KO-LBM theory can be obtained 
from equilibrium measurements, a clear comparison with the theory would
seem to be possible. However, as BC have discussed, their experiments violated
an almost universal theoretical assumption for this class of
 nonequilibrium phenomena, namely, that the
temperature is a control parameter. Rather, the quenches occurred by rapidly
decreasing the pressure and then holding it constant during the
decomposition. On the timescale of their experiments, no heat from the container
walls was able to reach the portion of the liquid being probed; thus
the decomposition occurred adiabatically rather than isothermally.

The problem with controlling only the pressure is that unmixing releases
 heat (being exothermic for most simple liquids), which causes the temperature
of the sample to increase with time. Further, fluid motion is enhanced during decomposition.
Theories of dynamic critical phenomena and the KO theory itself predict 
that the characteristic relaxation time of a binary liquid
scales as $\xi^{3+z_\eta} \sim \epsabs^{-1.94}$, where
$\epsilon = {T/T_c -1}$ is the reduced temperature, with $T$ and $T_c$ being
the absolute and critical temperature, 
respectively.\footnote{The exponent $z_\eta$
arises from the temperature dependence of the shear viscosity. See 
Table \ref{table3MPNEData}.} 
 The experiments of BC were done at reduced temperatures 
around $10^{-5}$, and so small changes in $T$ could cause large changes
in the relaxation time, making comparison with theory potentially troublesome.
More fundamentally though, what does one mean by ``temperature'' when a system
is driven so far from equilibrium?

The kinetics of phase transformation under adiabatic conditions was first examined theoretically by Schmelzer and Ulbricht for nucleation.\cite{schmelzer89} Onuki later performed a more careful analysis appropriate to near the gas-liquid critical point.\cite{onuki96}
Schmelzer et al.\ extended their earlier study to decomposition.\cite{schmelzer91,milchev94} This research focused on systems dominated by single particle diffusion, such as metals, far away from any critical point, in the approximation of uniform temperature changes only.
More recent work has examined the effects of coupling non-uniform temperature changes to the local concentration field during decomposition.\cite{lebedev19}
There has also been research in biology regarding kinetic processes under adiabatic conditions.\cite{mosgaard13,thoke18}

This paper has two purposes. First, it examines decomposition under conditions appropriate to the experiments of BC in critical binary liquids. It goes beyond the mean-field Cahn-Hillard approach of Schmelzer et al.\ by generalizing the stochastic theory of KO/LBM appropriate to binary liquids, but also by considering the parition of the degrees of freedom into slow and fast modes and the relevance of that to the meaning of temperature. Second, it makes a quantitative comparison, with no adjustable parameters, with the experiments of BC. A letter describing this work has been published.\cite{donley93}. The present paper describes in detail the theory, and also explores some necessary concepts not covered in the letter.

\section{\label{sec:adiadecomp}Adiabatic Decomposition}
In this section a theory of adiabatic decomposition in a binary substance
is presented. The theory generalizes any isothermal, statistical theory
of decomposition, such as KO or LBM.

In what follows, the equilibrium properties of critical binary fluids
 will be used to justify some theoretical approximations. 
Table \ref{table3MPNEData} below contains equilibrium data of 3MP+NE relevant
to the theory here. Table \ref{tableExpsAmps} below contains relevant 
critical exponent values and amplitude relations. In this work, the critical
point for a given pressure $P$ is denoted by the
concentration $c_c$ and temperature $T_c$. As mentioned above, the reduced
temperature $\epsilon = T/T_c-1$.  
For 3MP+NE at the pressures of interest, $T_c$ varies linearly with
pressure, so $dT_c/dP$ is a constant.\cite{clerke83}
In the critical region, the miscibility gap has the scaling form $\Delta
c=2B\epsabs^{\beta}$; the correlation length 
$\xi=\xi^{\pm}_0\epsabs^{-\nu}$; and the 
susceptibility $\chi=\Gamma^{\pm}\epsabs^{-\gamma}$, with
$\beta, \gamma$ and $\nu$ being critical exponents and $B, \xi_0^\pm$ and
$\Gamma^\pm$ being critical amplitudes.
Here, ``+'' refers to a one-phase value obtained on the critical isobar
above $T_c$, while ``-'' refers to a two-phase coexistence value below $T_c$.  
The quantities $c_c,B,\xi_0,\Gamma$ and any other critical amplitude 
mentioned in this text have either been shown experimentally 
to be, or are assumed to be constant over the pressures of 
experimental interest.\cite{clerke83}
\begin{table}[b]
\caption{\label{table3MPNEData}
Equilibrium one-phase ($T>T_c$) data of an on-critical mixture of
3MP+NE relevant to the present work. All units are MKS.}
\begin{ruledtabular}
\begin{tabular}{llr}
\textrm{Parameter} & & \textrm{Ref.}\\
\colrule
\textrm{Critical Temperature} & $T_c=300\,^{\circ}{\rm K}$ & \cite{clerke83}\\
& ${dT_c/dP} = 3.497\times10^{-8}\, {\rm K/Pa}$ & \cite{baileyThesis}\\
& &\\
\textrm{Critical Mass Density} & $\rho_c=7.92\times 10^2\, {\rm kg/m^3}$ &
\cite{greer73}\\
& & \\
\textrm{Correlation Length} & $\xi = \xi_0^+\epsabs^{-\nu}$ & \\
& $\xi_0^+ = 2.207\times 10^{-10}\, {\rm m}$ & \cite{baileyThesis} \\
& \textrm{with} $\nu = 0.632$ & \\
& & \\
\textrm{Hydrodynamic Shear} & $\eta_s={\bar\eta}(Q_0\xi)^{z_\eta}$ &\\
\textrm{Viscosity}& ${\bar\eta}= 3.76\times 10^{-4}$ {\textrm Pa-sec} & \cite{sorenson82}\\
& $Q_0= 1.4\times 10^8\, {\rm m}^{-1}$ & \cite{burstyn83}\\
& $z_\eta = 0.063$ & \cite{burstyn83}\\
& & \\
\textrm{Isobaric Heat Capacity} & $C_p=C_0^+\epsabs^{-\alpha}+C_b$ &\\
& $C_0^+ = 2.881\times 10^2$ \textrm{J/(K-kg)} & \cite{baileyThesis}\\
& $C_b=1.7074\times 10^3$ \textrm{J/(K-kg)} & \cite{baileyThesis}\\
& \textrm{with} $\alpha=0.105$ &\\
& & \\
\textrm{Thermal Expansion} &$\alpha_p=A_0^+\epsabs^{-\alpha}+A_b$&\\
\textrm{Coefficient} & $A_0^+=2.6597\times 10^{-5}$ $\textrm{K}^{-1}$ & \cite{baileyThesis}\\
& $A_b = 1.284\times 10^{-3}$ $\textrm{K}^{-1}$ & \cite{baileyThesis}\\
& \textrm{with} $\alpha=0.105$&\\
& & \\
\textrm{Adiabatic Compressibility} & $K_s\approx 1.07\times 10^{-9}$ \textrm{Pa}& \cite{tanaka83}\\
\end{tabular}
\end{ruledtabular}
\end{table}

\begin{table}[b]
\caption{\label{tableExpsAmps}
Theoretical critical exponent and amplitude relations relevant to binary fluids in the present work.}
\begin{ruledtabular}
\begin{tabular}{ccc}
\textrm{Exponent} & \textrm{Value} & \textrm{Reference}\\
\colrule
$\nu$ & 0.632 & \cite{fisher85}\\
$\gamma$ & 1.2395 & \cite{fisher85}\\
$\alpha$ & 0.105 & \cite{fisher85}\\
\colrule
\textrm{Amplitude Ratio} & \textrm{Value} & \textrm{Reference}\\
\colrule
${\Gamma^+/\Gamma^-}$ & 4.95 & \cite{liu89}\\
${\xi_0^+/\xi_0^-}$ & 1.96 & \cite{liu89}\\
${C_0^+/C_0^-}$ & 0.523 & \cite{liu89}\\
${T_c A_0^\pm/(\rho_c C_0^\pm)}$ & ${dT_c/dP}$ & \cite{griffiths70}\\
${\alpha\rho_c C_0^+\Gamma^+/(k_B B^2)}$ & 0.0581 & \cite{liu89}\\
$\xi_0^+({\alpha\rho_c C_0^+/k_B})^{1/3}$ &  0.265 & \cite{liu89}\\
\end{tabular}
\end{ruledtabular}
\end{table}

\subsection{\label{subsec:basic_theory} Basic Theory}
The essence of the theory here is to exploit how one constructs 
the coarse-grained free energy $\calF$ used in previous theories of
decomposition.
It is defined as follows.\cite{langer74} Consider a binary mixture of $A$ and
$B$-type molecules in strong contact with an external reservoir at temperature
$T$. Let $c(\bfr)$ be the concentration of $A$-type
molecules in a cell of size $a^3$ centered at position $\bfr$.
The cell size is mesoscopic on the order of the equilibrium correlation
length $\xi$, which for mixtures in the critical region can be
hundreds or even thousands of angstroms. The coarse-grained free energy
$\calF\equiv\calF\conc$, is a functional of the concentration field $c(\bfr)$.
 In mean-field theories a change in $\calF\conc$ due 
to a change in the concentration at some point $\bfr$
acts as a local thermodynamic driving force or 
chemical potential $\mu(\bfr)$.  Gradients in $\mu(\bfr)$
in turn cause mass diffusion.  

The coarse-grained free energy is constructed by
fixing the value of $c(\bfr)$ in each cell and performing the partition sum 
over all states of the system consistent with the configuration
$\conc$. Let the microscopic Hamiltonian be $H$, then
\begin{equation}
exp\bigl(-{\calF\conc + F_r\over k_BT}\bigr)=\sum_{{\rm states}\atop 
{{\rm consistent}
\atop {\rm with}{ } \conc}}exp(-{H\over k_BT}),
\label{eq:3a}
\end{equation}
where $k_B$ is Boltzmann's constant.
The coarse-grained free energy $\calF$ then describes the properties
of all concentration modes of wavenumber $k$ less than some 
cut-off $\Lambda\sim 1/a$.
The quantity $F_r$ is the part of the total equilibrium free energy that
is independent of the configuration $\conc$. For example, it is assumed that
$F_r$ contains all vibrational degrees of freedom, which
give the dominant contribution to a liquid's entropy.
In addition, the short wavelength concentration modes ($k>\Lambda$)
 contribute to $F_r$, but they also contribute to $\calF$ by 
renormalizing its coefficients. Because the long wavelength 
concentration modes don't contribute to $F_r$, $F_r$ is an 
analytic function of $T-T_c.$\cite{ma76}

Since the interest here is in the kinetics of phase separation, 
in integrating out these degrees of freedom it is assumed that they
relax very quickly compared to the modes described by $\conc$.  
That is, the modes in $F_r$ are able to equilibrate between any
characteristic change in $\conc$. Note that the strong contact 
with an external bath appears to make possible the
separation of the partial free energy into two terms $\calF$ and
 $F_r$. However, what if there were not strong contact?

\subsubsection{\label{subsubsec:closed_system} Closed System}
To examine the case of poor or no contact with an external bath, it is
helpful to look first at a system that is closed, i.e. has a fixed total energy $E_t$ and fixed volume $V$. This is also the case examined by Schmelzer and Milchev.\cite{schmelzer91}

Now, will non-uniformities in the temperature $T$ and pressure $P$ occur as the closed system evolves? For solids, the answer is yes for temperature and for fluids the answer is yes for both. For binary fluids, Mountain and Deutch examined the coupling of concentration, temperature and pressure fluctuations in equilibrium. They developed a way to estimate if this coupling is appreciable for any given simple liquid.\cite{mountain69} Here, the standard approximation made in the theory of dynamic critical phenomena will be taken and this coupling will just be assumed to be small.\cite{hohenberg77} The focus then is on uniform temperature and pressure changes caused by unmixing and fluid motion, which themselves turn out to be appreciable for 3MP+NE and indeed alter the unmixing kinetics.

For the closed system, the relevant ``free energy'' is the
entropy $S$.  In analogy with the definition for $\calF$ and 
following Boltzmann, define
\begin{eqnarray}
S(\conc,E_t)= k_B\ {\rm ln}\Biggl [
\sum_{{\rm states}\atop 
{{\rm consistent}
\atop {\rm with}{ } \conc}}\delta (H-E_t)\Biggr ],
\label{3b}
\end{eqnarray}
as the entropy of a system with a fixed
configuration $\conc$ and total energy $E_t$. Here, $\delta(x)$
is the Dirac ``delta'' function at point $x$.  
 
However, for the approximations to follow the interest is still
with $\calF$. To construct $\calF$, define
the partition function
\begin{eqnarray}Z=&&\sum_{{\rm states}\atop {{\rm consistent}
\atop {\rm with}\ \conc}} exp(-\beta H)\nonumber\\
 =&&\int\ dE^\prime\  exp(-\beta E^\prime)\Gamma(E^\prime,\conc),
 \label{3c}
 \end{eqnarray}
where $\beta$ is a parameter to be determined, and 
$\Gamma(\conc,E^\prime)=exp(S(\conc,E^\prime)/k_B)$ is the number
of accessible states of the system with an energy $E^\prime$
and coarse-grained configuration $\conc$.  Expanding $S$ about
$E_t$ it is found that, for the case in which $\Gamma$ is a 
macroscopic  number, $S$ is related to $Z$ by
\begin{equation}S(\conc,E_t)-k_B\beta E_t= k_B{\rm ln}\\ Z,
\label{3d}\end{equation}
with
\begin{equation}k_B\beta=\parsparet.\label{3e}\end{equation}

Now, one can write $S$ as the sum of two terms:
\begin{equation}S(\conc,E_t) =\calS\conc+S_r,
\label{3f}\end{equation}
where $S_r$ is the part of the entropy independent of the configuration $\conc$. Let the energy associated with $S_r$ and $\calS\conc$ be $E_r$ and $\calE\conc$, respectively.
Then,
\begin{equation}
E_t=\calE\conc + E_r.\label{3g}
\end{equation}
If the concentration field $\conc$ is held fixed, then so will be
$\calS\conc$ and $\calE\conc$; thus,
\begin{equation}
\parsparet=\parsparev\equiv{1\over T_r}.
\label{3g2}
\end{equation}
Combining Eqs. (\ref{3c})-(\ref{3g2}) then gives
\begin{equation}exp\bigl(-{\calF\conc + F_r\over k_B T_r}\bigr)=
\sum_{{\rm states}\atop 
{{\rm consistent}
\atop {\rm with}{ } \conc}}exp(-{H\over k_B T_r}),
\label{eq:3h}\end{equation}
where the coarse-grained free energy
\begin{equation}\calF=\calE-T_r\calS,\label{3i}\end{equation}
and
\begin{equation}
F_r=E_r-T_r S_r.\label{eq:3j}
\end{equation}

Comparing Eqs. (\ref{eq:3a}) and (\ref{eq:3h}) it can be seen for this
closed system that the degrees of freedom contributing to $F_r$
 act explicitly as a 
reservoir for the concentration modes described by $\calF$. 
For the remainder of this work, the degrees of freedom
that contribute to $F_r$ will be called the reservoir, and those that are
described by the concentration field $\conc$ and contribute to $\calF\conc$
will be called the slow modes.

For this closed system, the equilibration process is as follows. The system is
prepared in some nonequilibrium state with reservoir energy $E_r$ 
and coarse-grained configuration $\conc$ (and thus energy $\calE\conc
$).  The system is then released and the configuration $\conc$ evolves.
The evolution is driven by $\calF\conc$, which is a function of the reservoir 
temperature $T_r$.  As the $\conc$ changes, the coarse-grained
 energy $\calE\conc$ changes, and, because the total energy is
constant, the reservoir energy $E_r$ changes.  A change in $E_r$ 
implies a change in $T_r$, and this change in $T_r$ in turn 
affects the evolution of the $\conc$ and so on.  The system
eventually settles down into a state that maximizes the total
entropy. 

With the ideas above, an equation of motion for the system probability density, $\rho(\conc,t)$, for the case in which mass transfer is dominated by single particle diffusion (the solid model\cite{langer71}), can be derived.\cite{donleyUnpub}
But the coupling between the reservoir and $\cgmodes$ through the total energy $E_t$ in Eq.(\ref{3g}) makes the reservoir temperature $T_r$ an implicit function of the concentration field $\conc$. This equation is then not very useful.

However, as stated above, the primary concern is with average temperature changes associated with the decomposition process, so a simpler approach can be made. If $\rho(\conc,t)$ exists, then an average slow mode energy,
 $\ecgavg$, can be computed. For this average slow mode energy, there is a corresponding reservoir energy, $E_r = E_t - \ecgavg$. An average reservoir temperature, ``$\langle T_r\rangle$", can then be defined. Let this
average temperature then play the role of a pseudo control parameter,
let the thermodynamic driving force be $\calF=\calE-\langle T_r\rangle\calS$,
 and derive a separate equation of motion for $\langle T_r\rangle$.  In this way the same equation 
for $\rho(\conc,t)$ that has been used to describe
isothermal decomposition will be used, but any temperature dependent
parameter will now vary in time.
 It can be shown in equilibrium that the reservoir temperature is equal to the average system temperature
$({\partial E_t/\partial S_t})_V$, where
$S_t$ is the total equilibrium entropy.\cite{donleyThesis} When the system
is out of equilibrium, the relation between the reservoir and
 system temperatures is unclear - assuming the latter can
even be defined in a consistent manner. However, as will be seen below, such a
 relation is not necessary to determine the time evolution of the
 concentration field.

For the remainder of this article, let $T$ denote this average reservoir temperature,
i.e., $T = 1/(\partial S_r/\partial E_r)_V$, with $E_r\approx E_t - \ecgavg$. The time evolution of $T$ can be obtained for the closed system using the differential form of this equation:
\begin{equation}
dE_t=0\approx dE_r +d\ecgavg.
\label{3l}
\end{equation}
The reservoir energy is an equilibrium thermodynamic function so
\begin{equation}dE_r=\cvr dT,\label{3m}\end{equation}
where $\cvr = \parepartvInline$ is the reservoir heat capacity at constant
volume. Also,
\begin{equation}d\ecgavg=
 \decgdt dt,
\label{3n}\end{equation}
where the partial time derivative of the average is defined by
\begin{equation}
\decgdt \equiv \int {\cal D}c\ \calE\conc{\partial \rho(\conc,t)\over \partial t},
\label{decgdteq}
\end{equation}
with $\int{\cal D}c$ denoting an integral over the space of possible
concentration fields. Note that the term
$\bigl\langle{\partial \calE\conc/\partial T}\bigr\rangle dT$ does
not appear in Eq.(\ref{3n}) since in the above construction of $\calF\conc$,
$\calE\conc$ is independent of temperature. Combining
Eqs. (\ref{3l}-\ref{3n}) gives:
\begin{equation}{dT\over dt}=-{1\over
\cvr}{\partial\ecgavg\over\partial t}.
\label{3o}\end{equation}

Since the free energies are additive,
\begin{equation}
\cvr=\cv-\cvcg,
\label{3p}
\end{equation}
where $\cv$ is the equilibrium heat capacity of the total 
system, and $\cvcg$ is the contribution to $\cv$ from
the slow modes.  $\cv$ can be obtained from experiment and
$\cvcg$ can be calculated once $\calF\conc$ is defined:
\begin{equation}
\cvcg = -T\biggl ({\partial^2 \Fcg\over \partial T^2}\biggr )_V,
\label{cvcgeq}
\end{equation}
where $\Fcg = -k_BT\ {\rm ln} Z_{\rm sm}$ is the portion of the total system Helmholtz free energy from the slow modes, with the partition
function $Z_{\rm sm} = \int {\cal D}c\ exp(-{\calF\conc/ k_BT})$.

Eq.(\ref{3o}) will lead to a temperature change similar to that of Eq.(22) in Ref.\cite{schmelzer91} but with an important difference: the heat capacity in Eq. (\ref{3o}) is not for the entire system, but just the fast modes in the reservoir. Since the \cgmodes are the source of the singularity of the heat capacity at the critical point, the reservoir heat capacity can be much smaller than the total heat capacity near there, causing the temperature change to be larger than that predicted in Ref.\cite{schmelzer91}.

\subsubsection{\label{subsubsec:adiabatic_system} Adiabatic System}
These same ideas will now be applied to adiabatic decomposition.
In this case there is no heat flow between the system and the outside
world, and the pressure $P$ instead of the energy $E$ will be a control parameter.
Under these conditions a system undergoing phase separation will reach an
equilibrium state that minimizes its enthalpy $H_e=E+PV$.

However, as shown above, the Helmholtz free energy $F=E-TS$ seems to be the natural one to describe the decomposition process theoretically.
But if $P$ is controlled, the system volume $V$ an fluctuate. The above construction of $\calF$ does not handle well changes
in volume. With the coarse-grained cell size $a$ fixed, the 
number of cells changes as the volume is changed. However, 
for a typical pressure change $\Delta P$ 
for the quenches in the binary liquids of interest, the
fractional volume change ${\Delta V/V}\leq 10^{-5}\ll 1$.
So, given the level of  approximation of this theory, this 
ambiguity in the definition of the slow mode fields will be ignored.

As mentioned above, the only \cgmodes considered here will be those relevant to the isothermal case for critical dynamics in binary liquids, i.e., those described in Model H\cite{hohenberg77}: the concentration $c(\bfr,t)$ and fluid velocity $\bfv(\bfr,t)$ fields, so $\calF = \calF[c,\bfv]$.
Any effect of local fluctuations of the temperature and pressure are assumed to occur only in the reservoir, with the \cgmodes being influenced only by their spatially uniform values. The justification in decomposition for this is that average changes in $T$ and $P$ are expected to dominate during unmixing.

Given these approximations, what will be done
here is determine how the temperature and volume change in time during a pressure quench and subsequent decomposition. These time dependent values of $T$
and $V$ will then be inserted into the coefficients of the coarse-grained free energy $\calF$ to determine the evolution of the slow modes.

Since both the temperature and volume will change if the pressure changes, two
independent relations are needed to determined their time evolution.
The first relation is as follows.

To move the near-critical binary liquid from the one-phase region toward the
unstable portion of the two-phase region, the external pressure is dropped by
a differential amount $dP$, increasing the system (average) volume by
$dV$.  No heat is allowed to flow between the system and the external
world, so the average work done by the system on the external world (via
a piston, say) equals
the average change in the total system energy.
 Thus,
\begin{equation}dE_t = -PdV,
\label{3q}
\end{equation}
where $dE_t$ is given by Eq. (\ref{3l}) (though it is not
zero in this case obviously).

What, though, is the pressure $P$? In equilibrium,
\begin{equation}P = P_r + \Pcg,
\label{pressureeq}
\end{equation}
where $P_r = -({\partial F_r/\partial V})_T$ and  
$\Pcg = -({\partial \Fcg/\partial V})_T$ are the partial
pressures of the reservoir and slow modes, respectively, with
$\Fcg$ being defined below Eq.(\ref{cvcgeq}).
However, spinodal decomposition is a nonequilibrium process.
It necessarily does not allow the
 slow modes to relax completely during the quench. 
In the extreme case that the quench is so fast that the
slow modes are frozen, the contribution of these
modes to the pressure would be zero. Thus, the actual pressure of the
liquid on the container walls should be less than that given by Eq.(\ref{pressureeq}). On the other hand, it is assumed that the movement of the piston that
causes the drop in external pressure is 
slow enough so that the degrees of freedom in the reservoir are able to remain
in equilibrium. For example, any momentary density drop near the piston
wall is rapidly distributed throughout the liquid so no turbulence or
other inhomogeneous flow results.\cite{callen85} Thus, $P\geq P_r$.

Estimates of $\Pcg$, and the change of it, $\Delta \Pcg$, during the
 quench would be helpful here. As discussed above, upper bounds will 
be their equilibrium values.
In equilibrium, $\Pcg$ is a finite 
function of $\epsilon$ and so will change little for a near-critical
quench. So, an estimate of it at any point during the quench should be
sufficient. Now, the number of
of \cgmodes is $N= {V/a^3}\sim {V/\xi_f^3}$,
where $\xi_f$ is the equilibrium correlation length
at the final temperature $T_f$. Each slow mode will
have an energy of order $k_B T_f\sim k_BT_c$, and 
 so the equilibrium free energy of the \cgmodes is 
$\Fcg\sim-{k_B T_c V/\xi_f^3}$. In equilibrium then,
\begin{equation}
\Pcg\sim {k_B T_c\over \xi_f^3}.
\label{3x}\end{equation}
The short wavelength concentration modes also contribute
to $\calF$, so Eq. (\ref{3x}) may be an underestimate, 
but it should be accurate within an order of magnitude.
With $\epsfabs\sim 10^{-5}$ for the quenches of
3MP+NE of BC \cite{bailey93}, and using data from Table \ref{table3MPNEData},
it is found that $\Pcg \sim 1$ Pa in equilibrium. The
experiments of BC were done near standard pressure at sea level, which is
around $10^5$ Pa. Thus, 
\begin{equation}
{\Pcg\over P}\sim 10^{-5} \ll 1.
\label{3y}
\end{equation}
Also, the absolute change in pressure during a typical quench for the BC experiments
 (e.g., $\epsilon_i = 10^{-5}$ to $\epsilon_f = -\epsilon_i$) was
 $\vert\Delta P\vert\simeq 10^4$ Pa. An upper bound on $\vert\Delta \Pcg\vert$ is $\Pcg$, so
\begin{equation}
{\vert\Delta \Pcg\vert\over \vert\Delta P\vert} \leq 10^{-4} \ll 1.
\label{3y2}
\end{equation}
Thus, $P\simeq P_r$ throughout the quench and decomposition process.

Eq.(\ref{3q}) is completed by obtaining expressions for $dE_r$ and $d\ecgvavg$.
For this adiabatic case both the reservoir temperature and volume change, so
\begin{equation}dE_r=\parepartv dT+\pareparvt dV.
\label{3r}
\end{equation}
Likewise, the change in the average coarse-grained energy is
\begin{equation}
d\ecgvavg = {\partial\ecgvavg\over\partial V}dV +
{\partial \ecgvavg\over \partial t} dt.
\label{3r2}
\end{equation}
It will be seen in the next section that
${\partial\ecgvavg/\partial V}\simeq\ecgvavg/V$.
As discussed above, the relative volume changes for the near-critical quenches
with 3MP+NE were very small, so this term can be ignored.
Combining eqs.(\ref{3q}), (\ref{3r}) and (\ref{3r2}) gives a single
 equation relating $dT$ and $dV$ to $dP$ (i.e., $dP_r$).

A second relation is that of the differential change in the
reservoir pressure $P_r$ to changes in the reservoir temperature and volume:
\begin{equation}dP_r=\parprpartv dT+\parprparvt dV.
\label{3s}
\end{equation}

Combining Eqs. (\ref{3q}) and (\ref{3r})-(\ref{3s}), 
then using standard thermodynamic relations \cite{stanley71,
callen85}, and letting $dP_r/dt\approx dP/dt$, gives 
\begin{equation}\dtdt=\partparpsr{dP\over dt}-{1\over\cpr}\decgvdt,
\label{dTdteq}\end{equation}
and
\begin{equation}{dV\over dt}= -V \kapsr {dP\over dt}- {1\over T}\partparpsr
\decgvdt,
\label{dVdteq}\end{equation}
where
\begin{equation}
\partparpsr = {VT\alfpr\over\cpr}.
\label{dTdPrS}\end{equation}
Here, $\alfpr = \kaptr({\partial P_r/\partial T})_V={1/V}({\partial V/\partial T})_{P_r}$ is the reservoir isobaric thermal
expansion coefficient,  $\cpr =\cvr {\kaptr/\kapsr}$
 is the reservoir isobaric heat capacity,
$\kaptr=-{1/V}({\partial V/\partial P_r})_{T}$ is 
the reservoir isothermal compressibility, and
 $\kapsr=
-{1/V}({\partial V/\partial P_r})_{S_r}$ is the 
reservoir adiabatic compressibility.

The meaning of the above equations is this:
The pressure is changed at a known rate
$dP/dt$ and work is done on the system.  In the above
approximation all the work is done on the reservoir (
$\partparpsrInline$ and $\kapsr$ are reservoir functions).
 The reservoir reacts instantaneously and the temperature
$T$ and volume $V$ change at rates given by the first terms
in Eqs. (\ref{dTdteq}) and \ref{dVdteq}).  The coupling between the work
source and the \cgmodes is indirect.  As the change
in the reservoir causes $T$ and $V$ to change, the change in
$T$ and $V$ causes the coefficients in $\calF\conc$ to change.
A change in $\calF\conc$ causes the \cgmodes to be out of
equilibrium.  These modes then relax by exchanging energy
with the reservoir at constrained pressure, causing
$T$ and $V$ to change at a rate given by the second terms in
Eqs. (\ref{dTdteq} and \ref{dVdteq}). 

Looking more closely, consider a quench from the one-phase to the two-phase
region.  For simplicity, assume that the quench is very fast
so that all the \cgmodes will be frozen during it.
 Then during the quench the temperature will change at a rate given by the first term in Eq.(\ref{dTdteq}).  The final temperature $T_f$ is estimated (perhaps roughly) using
the full system thermodynamic function $\dtdpsInline$,
 where $S$ is the total entropy.\cite{clerke83}
Now, the total isobaric heat capacity, $\cp=\cv{\kapt/\kaps}$, where
$\kapt$ and $\kaps$ are the total isothermal and adiabatic compressibility,
respectively. Substituting this relation and that of $\cpr$ above into
 Eq.(\ref{3p}) gives:
\begin{equation}\cpr = {\kaptr\over\kapsr}{\kaps\over \kapt}C_p -
{\kaptr\over\kapsr}\cvcg.
\label{3ee}\end{equation}
However, since the contribution to $P$ from
the \cgmodes has been neglected, $\kaps\simeq\kapsr$ and 
$\kapt\simeq\kaptr$. Further, $\kapt\simeq \kaps$ at the temperatures
of experimental interest. 
Thus, 
\begin{equation}\cpr\simeq \cp - \Ccg,
\label{3ff}\end{equation}
where $\Ccg$ means either $\cvcg$ or $\cpcg$.
Further, from Tables \ref{table3MPNEData} and \ref{tableExpsAmps} it can be
 seen that the singular part
of the thermal expansion coefficient $\alfp$  is 
much smaller than the background part for the quenches
we are considering.  The \cgmodes contribute only (or almost
only) to the singular part of $\alfp$, which is much smaller than the 
background part for 3MP+NE; thus, $\alfpr\simeq\alfp$. So, since
  $\partparpsrInline\simeq {VT\alfp/(\cp -\Ccg)}> {VT\alfp/\cp}=\dtdpsInline$ , the 
temperature undershoots $T_f$ and reaches a value $T_{min}$.
Now after the quench the \cgmodes will relax and
the temperature will change at a rate given by the second term of Eq.(\ref{dTdteq}). As the system phase separates ${\partial\langle\calE\rangle/\partial t}<0$ and so the
temperature will increase, reaching its final value $T_f$ over time. On the other hand, since $\kapsr$ and $\partparpsrInline$ are both positive quantities, $V$ increases monotonically to its final value $V_f$.  As
$t\rightarrow\infty$ the system reaches a state of minimum enthalpy.

To complete the theory, useful expressions for $\alfpr,\cpr$, $\calF$ and
 $\calE$ are needed.
First, consider $\cpr$ which is given by Eq.(\ref{3ff}).  
Since $\cp$ is known, one could presumably determine $\cpr$ by calculating 
$\Ccg \simeq \cvcg$ using Eq.(\ref{cvcgeq}). Note though that if $\ecgvavg$ is known then so is $\Ccg$. In the next section, an explicit expression for $\ecgvavg$will be given, which can evaluated using equilibrium expressions of the functions obtained from the KO/LBM theory. The slow mode energy will be further approximated to depend only on the reduced temperature $\epsilon$. So,
\begin{equation}
\Ccg \approx {1\over T_c}\Bigl ({\partial \ecgvavg\over\partial\epsilon}\Bigr),
\label{32f}
\end{equation}
and the partial derivative implies holding fixed all parameters but $\epsilon$.
Now, since $\Ccg$ is expected to contain the singular piece of $C_p$, $\cpr$ will be an analytic function of $\epsilon$, which varies slowly around $\epsilon=0$. As such,
$\cpr$ will be further simplified by approximating it as its average value over the interval $\{\epsilon_i,\epsilon_f\}$, where $\epsilon_i$ is the
initial value of $\epsilon$.
With Eq.(\ref{32f}), averaging Eq. (\ref{3ff}) over $\epsilon$ from the initial
to the final temperatures, $\epsilon_i$ and $\epsilon_f$, respectively, gives:
\begin{eqnarray}
\cpr \approx && {1\over \Delta\epsilon} 
\int_{\epsilon_i}^{\epsilon_f}d\epsilon \bigl [ \cp - \Ccg \bigr ]\label{32h}\\
=&&{1\over \Delta\epsilon}\Bigl [
\int_{\epsilon_i}^{\epsilon_f}d\epsilon\ \cp -
 {1\over T_c}\bigl (\ecgvavg_f-\ecgvavg_i\bigr ) \Bigr ],\nonumber
\end{eqnarray}
where $\Delta\epsilon = \epsilon_f - \epsilon_i$. 
This approximation should be sufficient as long as $\epsilon_i\sim -\epsilon_f$,
that is, the quenches are neither too deep nor too shallow.

To calculate the equilibrium function $\alfpr$, consider a slow,
differential change in the pressure $P$ that allows the system
to remain in equilibrium. Since this process is reversible, the entropy
will remain constant.  Under these conditions Eq.(\ref{dTdteq}) can
be written as
\begin{equation}1= \Bigl\{{V\alfpr\over \cpr}
-{1\over T_c}{dT_c\over dP}\Bigr\}\biggl({\partial P\over \partial\epsilon
}\biggr)_S-{1\over T_c \cpr}\Bigl({\partial\ecgvavg\over\partial
\epsilon}\Bigr),
\label{32i}\end{equation}
where
\begin{equation}
\biggl({\partial P\over\partial \epsilon}\biggr)_S= 
T_c\Bigl[ \dtdps-\dTcdp\Bigr]^{-1}.\label{32j}
\end{equation}
 is an equilibrium function that relates changes in pressure
 to changes in the reduced temperature at constant entropy.
In terms of its components, $\dtdpsInline={VT\alfp/\cp}$. Combining  
 Eqs. (\ref{3ff}), (\ref{32f}), (\ref{32i}) and (\ref{32j}) give:
\begin{equation}\alfpr=\alfp- {\dTcdp\over VT_c} (\cp-\cpr).
\label{32n}\end{equation}
Using Tables \ref{table3MPNEData} and \ref{tableExpsAmps} it can be shown that
the singular parts of $\alfp$ and $\cp$ cancel in this equation.
Thus, with the approximation
above for $\cpr$, $\alfpr$ is a constant.

With the above equations, it is now possible to compute the final temperature
$T_f$ for a quench knowing the initial temperature $T_i$ and pressure change
$\Delta P$. Since decomposition is a nonequilibrium process, it is not expected
that $T_f$ will equal that computed using $\dtdpsInline$, which assumes constant entropy.

As the experiments are near $T_c$ it is useful to work with changes in
 $\epsilon$ rather than $T$. In terms of $\epsilon$, Eq.(\ref{dTdteq}) is:
\begin{equation}
{\cpr\over V}d\epsilon = \bigl (\alfpr - {\cpr\over VT_c}{dT_c\over dP}\bigr)dP - 
{1\over T_cV}d\ecgvavg.
\label{depsiloneq}
\end{equation}
Substituting Eq.(\ref{32h}) for $\cpr$, Eq.(\ref{32n}) for $\alfpr$, and then
integrating from the initial to final state gives
\begin{equation}
\int_{\epsilon_i}^{\epsilon_f}d\epsilon\ {\cp\over V} =
 \bigl(A_b - {C_b\over VT_c}{dT_c\over dP}\bigr)(P_f-P_i),
\label{epsilonfeq}
\end{equation} 
where $P_i$ and $P_f$ are the initial and final pressures, respectively. Also,
$A_b$ and $C_b$ are the background contributions to the total 
isobaric thermal expansion coefficient and isobaric heat capacity, respectively,
which are the same above and below $T_c$ (see Table \ref{table3MPNEData}, but
recognize that there the heat capacity is per unit mass).

Note that the average energy of the slow modes does not appear
in Eq.(\ref{epsilonfeq}). That is, $\ecgvavg$ determines the time evolution, including the
undershoot temperature $T_{min}$, but not the final temperature $T_f$. Also,
Eq.(\ref{epsilonfeq}) is the same equation that would be obtained by
assuming a reversible, constant entropy process.
In other words, the temperature rise produced during the
phase separation exactly compensates for the temperature undershoot caused by
the lack of equilibration of the slow modes during the quench. 

It can be shown that this expression for the final temperature $\epsilon_f$,
Eq.(\ref{epsilonfeq}), doesn't depend on the specific approximation, 
Eq.(\ref{32h}), for $\cpr$. At the least, Eq.(\ref{epsilonfeq}) will hold as
long as $\cpr$ is given by
 Eq.(\ref{3ff}) and the approximation for $\Ccg$ is consistent with the
value of $\ecgvavg$, which depends only on $\epsilon$ in equilibrium.

To understand better how this constant entropy approximation can be quantitatively accurate, it is helpful to recall the Rankine thermodynamic cycle, which is used to describe the operation of steam turbines.\cite{moran18} Like the Carnot and Brayton cycles, two of the four steps are isentropic.\cite{callen85}
One of these isentropic steps involves pressure quenching a superheated vapor by driving it through the turbine, causing it to rotate. The output product of this step is commonly a mixture of saturated liquid and gas,\cite{moran18} and so this step has realizations that are essentially identical to adiabatic decomposition.
No real steam turbine is ideal, yet efficiencies for this step can reach $90\%$.
Sources of inefficiency are friction, wear on the turbine blades caused by condensed droplets, any generated turbulence, and heat escaping to the outside. But the latent heat released during phase transformation of the vapor to liquid and gas does not affect the efficiency; it does not prevent the mixture from following a constant entropy path.

In the theory here, the cause of the constant entropy result is the use of properties of a binary
liquid in the critical region, which allows the neglect of the slow mode
pressure $\Pcg$ in and out of equilibrium. If $\Pcg$ were not small, then
the pressure response of the slow modes and thus the liquid would depend on the quench rate, so that
if the quench were fast the fluid entropy would increase in a manner similar to
that of a gas expanding into a vacuum.
A second reason is that, while the entropy of the slow modes is not
necessarily small (thus the reason for accounting for $\Ccg$), it containing the liquid latent heat, the dominant
 proportion of its change during phase separation is already accounted for
in an equilibrium, isentropic, process to get to the final state, as it is for the steam turbine. 
Thus, the overall exchange of energy between 
degrees of freedom in this nonequilibrium process is not much different
than if the process had been an equilibrium one, so whatever entropy
increase that does occur is small enough so that it can safely be approximated
as zero. Contrary to the original expectation then, the final temperature can be
computed accurately by just integrating Eq.\ (\ref{32j}).
\begin{figure}
\includegraphics[scale=0.31,trim= 0.0in 0.5in 0.0in 0.0in]{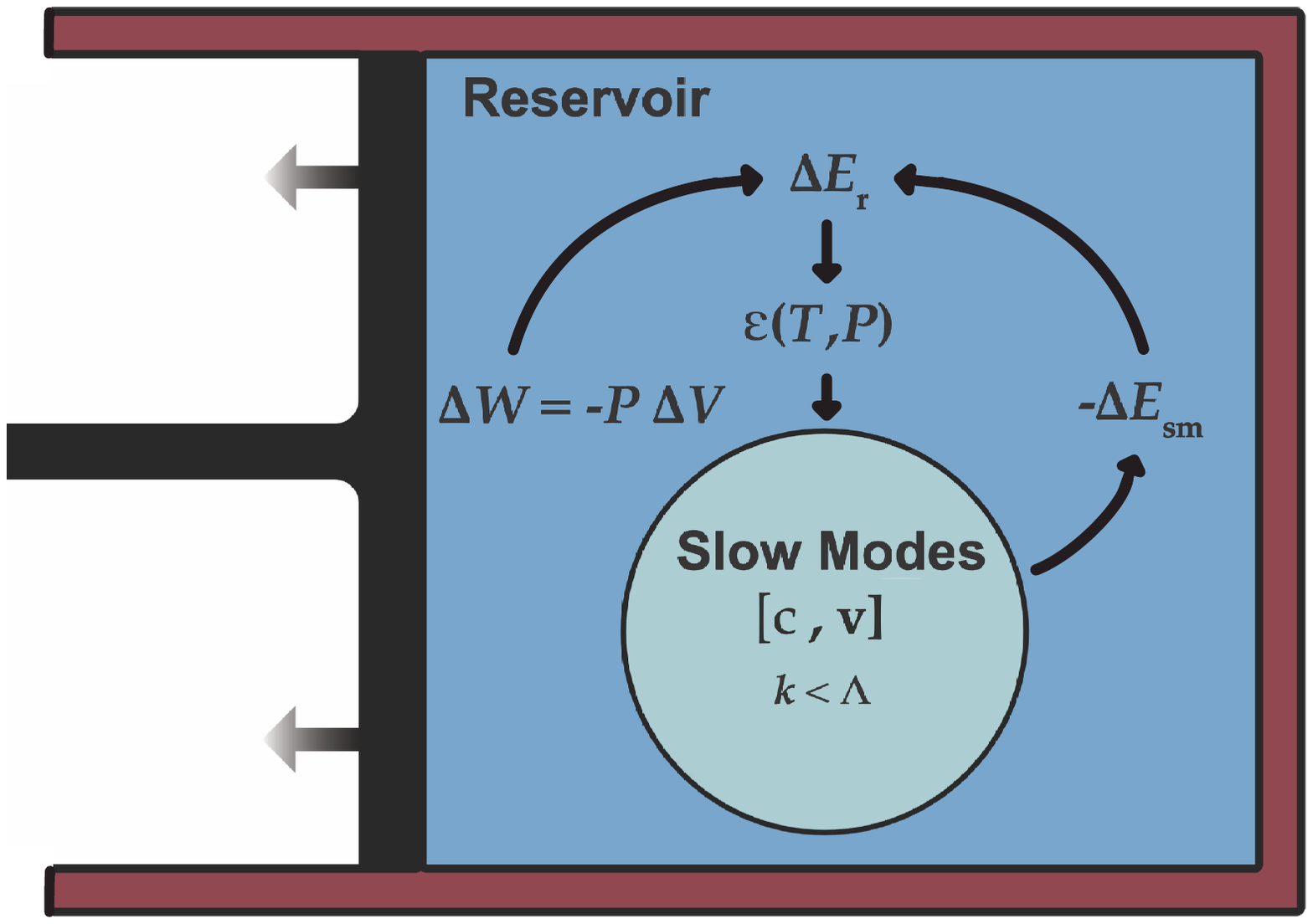}
\caption{\label{figAdiabaticDiagram}
Our theoretical model of adiabatic decomposition in critical binary liquids. The piston (black) expands the liquid, which is contained in an insulated container (brick red). All work by the piston is done only on the reservoir (blue) subsystem of the liquid. The reservoir temperature, $T$, and  pressure, equal to the external pressure, $P$, thus change. The slow mode (green-blue) subsystem of the liquid feels this change through the reduced temperature, $\epsilon(T,P)$. The slow modes then relax, changing their energy $E_{sm}\equiv\calEavg$, which is dumped back into the reservoir, causing a further change in $T$, etc.
}\end{figure}

\subsection{\label{subsec:free_energy}Temperature Dependent Coarse-Grained Free Energy}
The last elements of the adiabatic theory are expressions for the
temperature dependent coarse-grained free energy and energy.
 In isothermal decomposition, $\calF$ is taken to be that of Model H, which is a sum of a Cahn-Hilliard/Ginzburg-Landau free energy for the concentration field, $c(\bfr)$, and the kinetic energy of the fluid velocity field $\bfv(\bfr)$:
\begin{equation}\calF[u,\bfv] = \int_\Lambda d\bfr \Bigl [
{K\over 2}\bigl(\nabla u(\bfr)\bigr )^2 + f(u(\bfr)+c_0) + {1\over 2}\rho\bfv(\bfr)^2 \Bigr ],
\label{g33}\end{equation}
where $u(\bfr)\equiv c(\bfr) - c_0$ with $c_0$ being the average concentration. Also, $f(c)$ is the free energy
density of a uniform system at concentration $c$,
the gradient term is the lowest order correction to the free energy from
deviations of $u(\bfr)$ from zero\cite{cahn58}, and $\rho$ is the average mass density. 
In LBM and KO, the implicit cut-off is set to be
inversely proportional to the correlation length at the
quenched temperature, $T_f$, i.e.,  
$\Lambda\sim 1/\xi_f$.
Further, these theories also choose $f(c)$ to have the standard 
``$\varphi^4$'' form, it being the dominant correction to the quadratic term
in the critical region.\cite{ma76}

Given this, follow LBM and let
\begin{equation}
f(c)={k_BT_f f_1\over \xi_f^3}\phi(x),
\label{p3}\end{equation}
where 
\begin{equation}
\phi(x)={\zeta\over 2}x^2 + {\lambda_4\over 4}x^4.
\label{q3}\end{equation}
Here, $\zeta$ and $\lambda_4$ are constants to be determined.
Also, $x = {(c-c_c)/u_{sf}}$ is a reduced concentration, with 
$u_{sf} = B\epsfabs^\beta$ being half the miscibility gap at the 
quenched temperature $T_f$, so that the scaled free energy density
$\phi(x)$ is symmetric about the critical concentration.
Last, $f_1 = {\xi_f^3 u_{sf}^2/\chi_f}$, where $\xi_f$ and
$\chi_f$ are the correlation length and susceptibility, respectively, at $T_f$.
 In the critical region hyperscaling 
holds,\cite{stanley71} so
$f_1 = {(\xi_0^-)^3B^2/\Gamma^-}$, which is a temperature independent,
 dimensionless ratio of two-phase amplitudes.

Last, the gradient energy coefficient 
\begin{equation}
K=\lambda_K {k_BT_f\xi_f^2\over\chi_f},\label{eq:kappa}
\end{equation}
where $\chi_f = \Gamma^-\epsfabs^{-\gamma}$ is the susceptibility
at $T_f$, and $\lambda_K$ is a dimensionless number very close to unity.

How then should $\calF$ be generalized to describe kinetics in which
the temperature is not constant? While the
early-stage theories of KO and LBM can be used for computing equilibrium
states, they are not intended to describe properly static critical
phenomena. In spite of this, they do incorporate fluctuations to some
 degree. Thus, it can be expected that these fluctuations 
 will at least shift the apparent distance from the critical point, in a
manner similar to how they shift the coexistence concentrations
away from the minima of $f(c)$. So a correction 
for this shift in $T_c$ must be made in $\calF$.

What will be done here is just assume a simple temperature dependent form
for $\calF$ and compute its coefficients. Then, the free energy will be
 examined to determine
how well it predicts some equilibrium properties of a critical binary
mixture such as the equation of state and susceptibility. If the free energy
gives satisfactory results in regions important to the adiabatic decomposition
theory, then its form and the scheme used to compute it will be considered
adequate.

In that spirit, and given the arguments above (including those leading to
Eq.(\ref{3i})), assume
that the dominant temperature dependence in the theory is in $\zeta$ and 
$\lambda_4$. The latter changes to $\lambda_4{T/T_f}$, recognizing that the $x^4$ term arises from the configurational entropy. The $x^2$ term has both entropic and energetic origins, so let $\zeta$ be linear in $\epsilon$:
\begin{equation}
\zeta \rightarrow \zeta(\epsilon)\approx (\lambda_2 - \lambda_0){\epsilon\over\vert\epsilon_f\vert} - \lambda_0.
\label{zetaeq}\end{equation}
So, $\zeta(0) = -\lambda_0$ and $\zeta(\epsilon_f) = -\lambda_2$,
assuming $\epsilon_f < 0$.
Since the experiments are in the critical region, any other temperature
or volume dependence of $\calF$ will be ignored.

As $T_c$ itself changes when the density changes, $\epsilon$ will
be a function of both $T$ and $V$. However, for a system at constrained
pressure, $T_c$ is a function of pressure only, so $\epsilon$ will be
considered a function of $T$ and $P$ rather than $T$ and $V$.
In that manner, the equation of motion for $V$, Eq.(\ref{dVdteq}), will not
be used. 

With Eqs.(\ref{3i}) and (\ref{g33})-(\ref{zetaeq}), the average coarse-grained
energy $\calEavg = -\langle T^2\partial (\calF/T)/\partial T\rangle$ can be computed. In the Appendix it is shown that the heat of mixing dominates near the critical point, so
\begin{equation}
\calEavg\approx -(\lambda_2-\lambda_0) {1 \over \epsfabs}
{f_1\over 2}\langle x^2\rangle \biggl({k_BT_f V\over \xi_f^3}\biggr).
\label{eq:Ecgavg}
\end{equation} 
The one-point average $\langle x^2\rangle$ can be obtained
from the structure factor or one-point probability density, these
quantities being the subject of the next section.
The computation of the $\lambda_i$ parameters
will be described in Sec. \ref{sec:free_energy} below.

The theoretical model of adiabatic decomposition given here is summarized in Figure \ref{figAdiabaticDiagram}.

\section{\label{sec:hydrotheory}Kawasaki-Ohta Theory}
As mentioned above, the KO/LBM theory of early-stage decomposition will be
used to compare with experiment. This section describes the theory briefly.

KO is considered to be the most successful numerical
theory of decomposition in critical binary fluids. It, like LBM, is built upon
the theory of stochastic processes.\cite{vankampen81}
The KO theory consists of a
set of equations that describe the time evolution of the structure
factor $\shat (k,t)$, where $k\equiv \vert\bfk\vert$ is the wavevector. Contact with experiment is made by relating
$\shat (k,t)$ to the scattered radiation intensity
 $I(k,t)$.\cite{bailey94}

Now, let $u_\bfk$ be the Fourier transform of the concentration deviation 
$u({\bfr})$.  The structure factor $\shat(k)$ is
defined as 
\begin{equation}\shat(k)=\langle\vert u_\bfk\vert^2\rangle,
\label{h3}\end{equation}
and is the Fourier transform of the concentration-concentration
correlation function 
\begin{equation}S({\bfr}-{\bfr}_0)=\langle u({\bfr})u({\bfr}_0)\rangle.
\label{i3}\end{equation}
This function can be obtained from theory by taking 
moments of $\rho(\conu,t)$,
which, as mentioned above, is the probability density that the system is in a coarse-grained 
configuration $\conu$ at time $t$.

In KO theory, the time evolution of
the probability density $\rho(\conu,t)$ is determined by
a Fokker-Planck equation \cite{kawasaki78}:
\begin{equation}
{\partial\rho\over\partial t}=
\Bigl[ \calL_1+\calL_2\Bigr]\rho,
\label{fokkerplanckeq}\end{equation}
where the operators are given by
\begin{widetext}
\begin{equation}\calL_1=-\int d{\bfr}_1 d{\bfr}_2{\delta\over \delta 
u({\bfr}_1)}\nabla_1^2L_\Lambda({\bfr}_1-{\bfr}_2)
\Bigl[{\delta \calF\over\delta u({\bfr}_2)}+k_B T{\delta\over
\delta u({\bfr}_2)}\Bigr],
\label{L1operatoreq}\end{equation}
and
\begin{equation}\calL_2=\int d{\bfr}_1 d{\bfr}_2{\delta\over\delta
u({\bfr}_1)}\nabla_1 u({\bfr}_1)\cdot {\bf T}({\bfr}_1-
{\bfr}_2)\cdot\nabla_2 u({\bfr}_2)\Bigl[{\delta\calF\over
\delta u({\bfr}_2)}+k_B T{\delta\over\delta u({\bfr}_2)}
\Bigr].
\label{L2operatoreq}\end{equation}
\end{widetext}
Here, ${\delta/\delta u(\bfr)}$
is a functional derivative with respect to the concentration 
field at the point $\bfr$,
and ${\bf T}(\bfr)$ is the 
Oseen tensor with components $T_{\alpha\beta}=[\delta_{\alpha\beta}+
{\hat r}_\alpha{\hat r}_\beta]/(8\pi\eta_s r)$, with $\eta_s$ being the
hydrodynamic shear viscosity and ${\hat r}\equiv {\bfr/r}$. 

Eqs. (\ref{fokkerplanckeq}-\ref{L2operatoreq}) describe
phase separation driven by incompressible, overdamped fluid flow, the flow
in turn caused by gradients in the local chemical potential,
$\mu(\bfr) = {\delta\calF\conu/\delta u(\bfr)}$.
These equations are renormalized versions\cite{kawasaki77} of
bare stochastic equations\cite{kawasaki73}, which are
formally equivalent to the Langevin equations of Model H of
critical dynamics\cite{hohenberg77} in the overdamped approximation,
i.e., $\bfv = {\bf T}\cdot (\mu\nabla u + {\rm noise})$.

The operator $\calL_1$ results from integrating out concentration
fluctuations of wavenumber $k>\Lambda$.
The Onsager function $L_\Lambda(r)$ that appears in $\calL_1$ 
 is  weakly non-local and couples
these short-wavelength concentration modes, $\it via$ the fluid velocity
field, to the long-wavelength modes $k<\Lambda$.
$L_\Lambda(r)$ is the inverse Fourier transform of
\cite{kawasaki77}:
\begin{equation}\lhat(k_1)={1\over k_1^2}\int{d\bfk_2\over (2\pi)^3} 
\bfk_1\cdot {\bf\hat T}(\bfk_1-\bfk_2)\cdot \bfk_1\shat_{eq}(k_2).
\label{d3}\end{equation}
Here, the integral over $k_2$ runs from $\Lambda$ to an upper cut-off 
which is approximated as infinity.
 ${\bf\hat T}(\bfk)$ is the Fourier transform of the
Oseen tensor with components $[\delta_{\alpha
\beta} - {\hat k}_\alpha {\hat k}_\beta]/(\eta_s k^2)$.
$\shat_{eq}(k)$ is the equilibrium structure factor of 
mode $k\ (>\Lambda)$, and is taken to have a Lorentzian form:
\begin{equation}\shat_{eq}(k)={\chi\over 1+(k\xi)^2},
\label{shateq_lorentz}\end{equation}
where $\chi$ and $\xi$ are the susceptibility and correlation 
length, respectively, at temperature $\epsilon$ and concentration $c_0$.
 Clearly, as the temperature changes, $\shat_{eq}(k)$, and
therefore $\lhat(k)$, will be changing also. However, it can be shown
\cite{kawasaki77} that a reasonable approximation to Eq. (\ref{d3}) is 
\begin{equation}
\lhat\simeq{\chi\over 3\pi^2\xi^2 \eta_s\Lambda}.\label{g3}
\end{equation}
Making use of Tables \ref{table3MPNEData} and \ref{tableExpsAmps} for
 $\chi,\xi$ and $\eta_s$, it is found that $\lhat$ has
a weak temperature dependence.  Thus, if the quenches are relatively
fast, $\shat_{eq} (k)$ can be set to its value at the 
final equilibrium temperature $\epsilon_f$. Further, the weak temperature
dependence of the viscosity $\eta_s$ will also be ignored.  Then with 
these approximations, the only temperature dependence in (\ref{fokkerplanckeq}) 
appears in the coarse-grained free energy, $\calF[c,\bfv]$. 

An equation of motion for the structure factor $\shat(k,t)$ was
derived by KO using Eqs. (\ref{g33}) and (\ref{fokkerplanckeq}-\ref{d3}). 
 To evaluate two-point correlation functions in $\calL_1$ other than $S(r)$, they used the LBM self-consistent, first order expansion for the two-point probability density:
\begin{equation}\rho_2^{lbm}(u_1,u_2) = \rho_1(u_1)\rho_1(u_2)\Bigl [
 1 + {u_1 u_2\over \uu2^2}S(r_{12})\Bigr ],
\label{ansatz}\end{equation}
where $r_{12}\equiv\vert \bfr_1-\bfr_2\vert$, $u_1 \equiv u(\bfr_1)$, etc.
This approximation is expected to work best during
the early stage of decomposition when the growing domains are not much
larger than a few equilibrium correlation lengths and sharp interfaces
have not yet formed.\cite{langer75} Implicit in the LBM derivation is
an important constraint that averages taken with respect to $\rho_2$ must reduce to their exact form in the limit $r_{12}\rightarrow 0$. That is,
for arbitrary functions $h(u)$ and $g(u)$,
$\langle h(u_1)g(u_2)\rangle \rightarrow \langle h(u)g(u)\rangle$ as
$r_{12}\rightarrow 0$, where the latter average is
taken with respect to the one-point probability density $\rho_1(u)$.
Implementing this constraint in a simple way, and using the LBM approximation above, gives
an equation for $\rho_2$:
\begin{eqnarray}
\rho_2(u_1,u_2) \approx &&\rho_2^{lbm}(u_1,u_2) +a^3\delta(\bfr_1-\bfr_2)\Bigl [
 \label{ansatz2} \\
&&\rho_1(u_1)\delta(u_1-u_2) - \rho_2^{lbm}(u_1,u_2)\bigr \vert_{r_{12}\rightarrow 0}
\Bigr ].\nonumber\end{eqnarray}

The $\calL_2$ contribution to $\shat(k,t)$ contains a four-point correlation
function. KO argued that during the early stage of decomposition the 
coupling between modes in this correlation function would be close to gaussian.
In this approximation the four-point correlation function reduces to a product of two-point ones.\cite{kawasaki78}

The result is:
\begin{eqnarray}
{\partial \shat(k_1)\over\partial t}&&=-2\lhat(k_1)
k_1^2\Bigl[\bigl(Kk_1^2+A\bigr)\shat(k_1)-k_B T\Bigr]\label{dshatdteq}\\
+&&2\int^\Lambda{d\bfk_2\over (2\pi)^3}{\bfk_1}\cdot
{\bf\hat T}(\bfk_1-\bfk_2)\cdot{\bfk_1} \nonumber \\
\times \Bigl[K\bigl(&&k_2^2-
k_1^2\bigr )\shat(k_2)\shat(k_1)+k_B T\shat(k_2)-k_B T\shat(k_1)
\Bigr].\nonumber
\end{eqnarray}
Here,
\begin{equation}A = {1\over \uu2}\bigl\langle u{\part f(u+c_0)\over
\part u}\bigr\rangle,
\label{Aeq}\end{equation}
where the averages are taken with respect to $\rho_1(u)$.

It can be shown\cite{kawasaki78} that the operator $\calL_2$ doesn't
contribute directly to the equation of motion for 
$\rho_1(u)$.  Given this, the derivation of the 
equation of motion for $\rho_1(u,t)$ from Eq.(\ref{fokkerplanckeq}) is almost
identical to the one in LBM.  It is found:
\begin{equation}{\partial\rho_1(u)\over\partial t}={\partial\over\partial u}
\Bigl[G(u)\rho_1(u)+k_B T{L\over a^3}{\partial\rho_1(u)
\over\partial u}\Bigr ],
\label{l3}\end{equation}
where
\begin{equation}
G(u)=W{u\over\uu2}+L\Bigl[{\partial f\over\partial u}-
\bigl\langle{\partial f\over\partial u}\bigr \rangle- uA \Bigr ],
\label{m3}\end{equation}
\begin{equation}W=\int_0^\Lambda {dk\over 2\pi^2} k^4 \lhat(k)
\bigl(Kk^2+A)\shat(k),
\label{n3}
\end{equation}
and
\begin{equation}L=a^3\int_0^\Lambda {dk\over 2\pi^2} k^4 \lhat
(k).\label{o3}\end{equation}

For the initial conditions of these equations, the equilibrium solution of them
will be used. Setting the RHS of Eq.(\ref{l3}) to zero yields:
\begin{eqnarray}
\rho_{1eq}(u) &&= exp \Bigl[-{u^2\over 2\uu2} + b_0 \\
-&&{a^3\over k_BT}\bigl(f(u+c_0) - u\mu - {u^2\over 2}A\bigr)\Bigr] \nonumber
\label{rho1eq}
\end{eqnarray}
where $b_0$ is a normalization constant, and $\mu\equiv\langle{\partial f/\partial u}\rangle$ is the chemical potential. Both $\mu$ and $A$ are determined self-consistently, while $\uu2 = S(0)$ is an input obtained from the Fourier transform of the structure factor. The equilibrium structure factor is found by setting the RHS of Eq.(\ref{dshatdteq}) to zero, giving
\begin{equation}
\shat_{eq}(k) = {k_BT\over Kk^2 + A}.
\label{shateq}
\end{equation}

These kinetic and equilibrium equations were solved numerically.

\section{\label{sec:scale_and_num}Scaling and Numerical Solution}
\subsection{\label{subsec:scaling}Scaling of the Equations}
For numerical computation, it is helpful to scale the above equations.
While in adiabatic decomposition the temperature will necessarily be
changing with time after the quench, the final equilibrium temperature
will still be the relevant one. So, as in KO and LBM, the scaling will
be done with respect to system properties at $\epsilon_f$.

Define the scaled wavevector cut-off $\alfstar=\Lambda\xi_f$,
where $\alfstar$ is a number close to 1. 
Define also the dimensionless wavevector, $q=k\xi_f$; 
distance, ${\tilde r} = r/\xi_f$; 
structure factor, $\stil(q) = \shat(k)/\chi_f$; 
relative concentration, $y = u/u_{sf}$;
average concentration, $x_0 = {(c_0-c_c)/u_{sf}}$; 
and time, $\tau = k_BT/(6\pi\eta_s\xi_f^3)t$.
As for LBM, the cell volume
 $a^3 = (\int {d\bfk}/(2\pi^3))^{-1} = {6\pi^2\xi_f^3/(\alfstar)^3}$.

It is convenient to scale the Onsager function, Eq.(\ref{d3}), as 
\begin{eqnarray}
\sigma(q)=&&{6\pi \xi_f\eta_s\over\chi_f }
\lhat(q/\xi_f)\label{sigmaqeq}\\
=&&K(q)-{3\over 2\pi}
\int_0^{\alfstar}dm\  Q(q/m){1\over 1+m^2}
\nonumber\end{eqnarray}
where $K(q)$ is a Kawasaki function\cite{kawasaki78}:
\begin{equation}K(q)={3\over 4}\Bigl[({1\over q}-{1\over q^3})\arctan(q) 
+{1\over q^2}\Bigr],\label{kawasakifcneq}\end{equation}
and
\begin{equation}Q(x)={1\over 2}\Bigl[{1\over x}+ {1\over x^3}\Bigr]
{\rm ln}\bigl\vert {1+x\over 1-x}\bigr\vert-{1\over x^2}.
\label{41l}\end{equation}

Changing to the new scaled variables and performing any angular
integration, the equation of motion for the structure factor becomes:
\begin{eqnarray}
{\partial\stil(q)\over\partial\tau}&&=-2\sigma(q)q^2
\Bigl[(\lambda_Kq^2+{\tilde A})\stil(q)-1\Bigr]\label{dstilqdteq}\\
+&&{3\over \pi}q^2\int_0^{\alfstar}dm\ Q(q/m) \nonumber \\
\times\Bigl[&&\lambda_K(m^2-q^2)
\stil(q)\stil(m)+\stil(m)-\stil(q)\Bigr],\nonumber
\end{eqnarray}
where 
\begin{equation}
{\tilde A}={1\over \yy2}\bigl\langle y{\partial\phi(y+x_0)\over \partial y}
\bigr\rangle.\label{Atileq}\end{equation}

The kinetic equation for  the one-point probability density $\rho_1$ is now
\begin{equation}{\partial\rho_1(y)\over\partial\tau}=\omega
{\partial\over\partial y}\Bigl[g(y)\rho_1(y)
+{\partial\rho_1(y)\over\partial y}\Bigr],
\label{drho1ydteq}\end{equation}
where
\begin{equation}
\omega = {1\over f_1}\int_0^{\alfstar}{dq\over 2\pi^2}q^4\sigma(q),
\label{omegaeq}\end{equation}
and
\begin{equation}g(y)={{\tilde W}y\over \yy2}+f_0\Bigl[{\partial\phi\over\partial y}
-\langle{\partial\phi\over\partial y}\rangle-y{\tilde A}\Bigr],
\label{gyeq}\end{equation}
with
\begin{equation}
{\tilde W}={1\over f_1\omega}\int_0^{\alfstar}
{dq\over 2\pi^2}\ q^4\sigma(q)(\lambda_Kq^2+{\tilde A})\stil(q).
\label{wtildeeq}\end{equation}
Here, $f_0 = {6\pi^2 f_1/(\alfstar)^3}$.

Last, using Eq.(\ref{dTdteq}), it can be found that the equation of motion for
the reduced temperature is:
\begin{equation}
{d\epsilon\over d\tau} = \Bigl [ {\alfpr\over \rho_c\cpr} - {1\over T_c}\dTcdp
\Bigr] {dP\over d\tau} - {k_B\over\xi_f^3\rho_c\cpr}{d{\tilde e}\over d\tau},
\label{depsdteq}\end{equation}
where $\rho_c$ is the critical mass density and $\cpr$ is now the reservoir
heat capacity per unit mass, and the average coarse-grained energy,
Eq. (\ref{eq:Ecgavg}), properly scaled, is: 
\begin{eqnarray}
{\tilde e} =&&{\xi_f^3\over Vk_BT_f}\ecgvavg\nonumber\\
 =&&-(\lambda_2-\lambda_0){f_1\over2\epsfabs}\langle x^2\rangle.
\label{ecgavgeq}\end{eqnarray}

The scaled form of the equilibrium equations, (\ref{rho1eq}) and (\ref{shateq}), are, respectively,
\begin{eqnarray}
\rho_{1eq}(y) &&= exp\Bigl[ -{y^2\over 2\yy2} + b_0 \\
-&&f_0\bigl(\phi(y+x_0) - y\langle{\partial\phi\over\partial y}\rangle -{y^2\over 2}{\tilde A}\bigr) \Bigr],\nonumber
\label{rho1yeq}
\end{eqnarray}
and 
\begin{equation}
\stil_{eq}(q) = {1\over \lambda_Kq^2 + {\tilde A}},
\label{stileq}\end{equation}
where again $b_0$ is a normalization constant.

From these equations it can be seen that an (linear) isothermal quench
and subsequent decomposition is
completely specified by the quench time $\tau_{quench}$, the ratio 
of the initial to the final scaled temperature, 
$\epsilon_i/\epsilon_f$, and the average concentration $x_0$.
In addition to the properties of the particular fluid one wants
to study, an adiabatic quench is completely specified by these same
quantities plus the change in pressure, $\Delta P$. 
The predictions of the  theory are also somewhat dependent on the
value of the scaled cut-off $\alfstar$. However, the degree of this 
dependence
will be minimized by the method of computing the $\lambda_i$ parameters
in the coarse-grained free energy, discussed below.

\subsection{\label{subsec:numerical_solution}Numerical Solution}
The scaled adiabatic equations were solved numerically as follows.

$\stil(q)$ was solved on a grid of $N_q$ points
$q_i = i\Delta q$, $i=1,...,N_q$ in $q$-space, with spacing
$\Delta q = 2\alfstar/N_q$.
$\stil(q)$ was set to zero for all grid points $q_i > \alfstar$. The reason the
grid was extended in this manner was to be able to inverse Fourier transform
$\stil(q)$ if need be. The $q$ grid point number $N_q$ was set to $2^9=512$ for
 any run with a maximum time
$\tau_{max}\leq 10^3$, and was set to $2^{10}=1024$ for longer runs of
$10^3<\tau_{max}\leq 10^4$. 
Similarly, $\rho_1(y)$ was solved on a grid of $N_y$ equally spaced points
$y_i$, $i=1,..,N_y$ in $y$-space, with
$y_1 = y_{min}+\Delta y/2$, $y_2 = y_{min}+3/2\Delta y$, and
$y_{N_y} = y_{max}-\Delta y/2$, where $\Delta y=N_y/(y_{max}-y_{min})$ and
$y_{min} = -y_{max}$.  To ensure that $\rho_1(y)$ could model properly
behavior near the coexistence curve at $\epsilon_f$, $y_{max}$ was set to 2.5. 
Also, $N_y=120$ for all results shown in this work.
It was found that no result shown here changed appreciably if $N_q$ and $N_y$
were increased beyond the above values.

Now, the adiabatic theory consists of ODE's for
 $\stil(q,\tau)$ and $\epsilon(\tau)$, and
a PDE for $\rho_1(y,\tau)$. To simplify the computation,
the PDE for $\rho_1(y,\tau)$ was converted into a set of coupled ODE's,
using a simple finite difference scheme. Let $\rho_{1i}$ and $g_i$ be the values
of $\rho_1(y)$ and $g(y)$ at the $i$th grid point $y_i$. Then, Eq. (\ref{drho1ydteq}) becomes
\begin{eqnarray}
{d\rho_{1i}\over d\tau}&& =\omega
\Bigl[g_{i+1}\rho_{1(i+1)} - g_{i-1}\rho_{1(i-1)}\Bigr]/(2\Delta y) \nonumber \\
+&& \omega\Bigl[\rho_{1(i+1)}+ \rho_{1(i-1)} - 2\rho_{1i}\Bigr]/\Delta y^2,
\label{drhody_discrete}\end{eqnarray}
with the boundary conditions $\rho_{11} = \rho_{1N_y}= 0$. A total of $N = N_q/2 + N_y-2 + 1$ ODE's  result. The integration of
these in time was done using the Bulirsch-Stoer method.\cite{numrecipes} 
The structure factor equations are stiff in the sense that the relaxation of
the high-$q$ modes is much faster than the low-$q$ ones. The Bulirsch-Stoer
method is not usually used for solving such an ODE type. Given that,
initially the equations were also solved using a commercial package built
for solving stiff ODE's.\cite{NAG} It was found that both methods yielded the
same results.

In past work, the PDE for $\rho_1(y)$ was solved instead using the
much faster ``double gaussian'' method.\cite{langer75} This method was
 analyzed for on-critical quenches, $x_0=0$, and found to give
 essentially identical results to the finite difference method at early times. 
However, it over-estimated the phase separation at later times when the
wavevector $q_m$ of the peak of the structure factor was less than $0.3$. 
For the BC experiments, data was available out to times such that 
$q_m < 0.2$. As a consequence, this approximation was not used here.

For each timestep, $\rho_1$ was normalized to prevent accumulation of round-off errors. The first  moment, $\langle y\rangle$ of $\rho_1$ was monitored to ensure that it remained zero. 
The integral of $\stil(q)$ and $\rho_1$ were also
monitored to ensure that they gave the same result for $\langle y^2\rangle$.

The equilibrium equations,
 (\ref{rho1yeq}) and (\ref{stileq}), were solved using the same grids for
$q$ and $y$, though here $N_q$ was set to $2^{12}=4096$. Simple iteration was
 used for them.
The initial guess for $\stil_{eq}(q)$ was a scaled version of
 Eq.(\ref{shateq_lorentz}) at the relevant initial $\epsilon$ and $x_0$.
However, if the LBM solution of $\stil_{eq}(q)$ differed appreciably from
the known value, bootstrap, i.e., a previous guess at a nearby
 temperature and concentration, was used instead.

\section{\label{sec:free_energy}Computing the Coarse-Grained Free Energy}
The last ingredients of the theory are values for the $\lambda_i$
parameters in the coarse-grained free energy, 
defined in Sec. \ref{subsec:free_energy} above. Similar to LBM,
these parameters were determined by using $\calF$ to compute the
equilibrium structure factor and chemical potential
on coexistence at the final temperature $\epsilon_f$, and at the critical
point $\epsilon=0$. There are a number of ways to accomplish this task. The
one probably most accurate for the least amount of effort is to use the
LBM equilibrium solution for $\stil(q)$ and $\rho_1(y)$.\cite{billotet79}
This scheme was used here. It amounts to applying $N$ equilibrium
conditions to the LBM equations for the $N$ unknowns, and  
finding the solution of them using the Newton-Raphson method.\cite{numrecipes} 
Here, all derivatives required by this method were computed numerically.

For here and elsewhere in this work, the free energy amplitude $f_1$
was set to 0.210, consistent with the critical amplitude values in Table
\ref{tableExpsAmps}.

On coexistence, $\epsilon=\epsilon_f$ and $x_0=1$, the scaled equilibrium
structure factor $\stil(q) = 1/(1 + q^2)$. Two relations obtained from this
equation were $\stil(0)=1$ and
\begin{eqnarray}
\langle y^2\rangle 
=&&{1\over 2\pi^2f_1}\int_0^{\alfstar} dq q^2\stil(q)\nonumber\\
=&&{1\over 2\pi^2f_1}\bigl[\alfstar - \arctan(\alfstar)\bigr].
\label{y2avgeq}\end{eqnarray}
A third relation is that the exchange chemical potential must 
vanish on coexistence:
${\tilde\mu} = \langle{\partial\phi/\partial x}\rangle = 0$.
These three equations are sufficient to determine $\lambda_K$,
$\lambda_2$ and $\lambda_4$ for any cut-off $\alfstar$.
The last parameter $\lambda_0$ was determined by
requiring that $\stil(0) = \infty$, i.e., $\tilde A = 0$, at the critical
 point, $\epsilon=0$ and $x_0=0$. 
Values for the $\lambda_i$ parameters for various cut-offs are shown in
Table \ref{tablelambda} below. Note that the value of
 $\lambda_0$ does not depend on the
{\it form} of the temperature dependence of $\zeta$,
Eq.(\ref{zetaeq}), only that it reduce to $-\lambda_0$ at $\epsilon = 0$. 
\begin{table}[b]
\caption{\label{tablelambda}
Coarse-grained free energy coefficients $\lambda_i$ for various
wavevector cut-offs $\alfstar$.}
\begin{ruledtabular}
\begin{tabular}{lllll}
{$\alpha^*$} &
{$\lambda_K$} &
{$\lambda_0$} &
{$\lambda_2$} &
{$\lambda_4$}\\
\colrule
1.0 & 1.0 & 0.3203 & 0.6915 & 0.6020\\
1.4 & 1.0 & 0.5169 & 0.8734 & 0.6704\\
{$\pi/2$} & 1.0 & 0.6054 & 0.9533 & 0.6913\\
\end{tabular}
\end{ruledtabular}
\end{table}

To justify some approximations made previously, it is helpful to 
examine the predictions of the equilibrium LBM theory on-critical above
$T_c$ and on-coexistence below it. 

\begin{figure}
\includegraphics[scale=0.33,trim= 0.4in 0.5in 0.4in 0.2in]{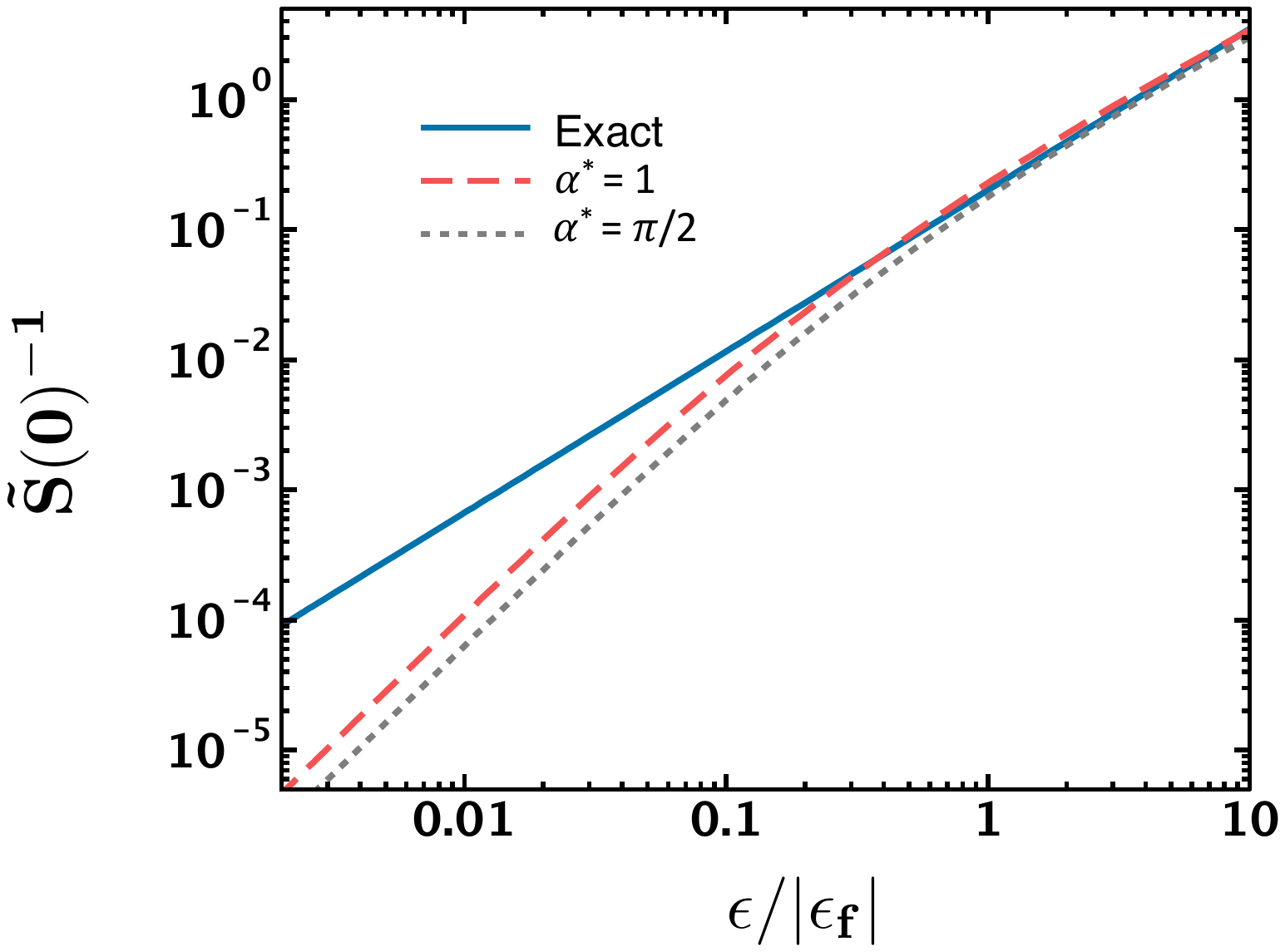}
\caption{\label{figchivseps}
LBM predictions, using the temperature dependent coarse-grained free energy
 defined in Sec. \ref{subsec:free_energy}, for the equilibrium, one-phase,
on-critical, inverse  susceptibility $\stil(0)^{-1}$ as a
function of the scaled temperature $\epsoverepsf>0$.
 Results for two cut-offs $\alfstar$ are shown, along with ``exact'' scaling
values.
}\end{figure}

Figure \ref{figchivseps} shows LBM predictions for the equilibrium,
on-critical inverse susceptibility $\stil^{-1}(0)$ as a function of
$\epsoverepsf>0$ for two cutoff values $\alfstar=1$ and $\pi/2$.
Also shown are the expected scaling predictions for a simple binary
 fluid (3-D Ising universality class):
$\stil(0)^{-1}=\Gamma^-/\Gamma^+\vert{\epsilon/\epsilon_f}\vert^{\gamma}$
where the amplitude ratio and exponent are obtained from
 Table \ref{tableExpsAmps}.
For $\epsoverepsf \ll 1$, it is found that LBM also predicts scaling
behavior, with the exponent $\gamma$ approximately equal to the
mean spherical model value of 2.\cite{stanley71}
Since the accepted 3-D Ising value of $\gamma\approx1.240$, the LBM theory does
 not perform
well in this limit as expected. However, it can seen for higher temperatures
the theory performs much better. For $0.5\leq\epsoverepsf\leq 10$, the
LBM predictions for $\stil(0)$ are within 10\% and 20\% of the exact
values for $\alfstar=1$ and $\alfstar=\pi/2$, respectively. For higher
temperatures, the agreement lessens, but $\stil(0)$ is small there 
anyways. It was found that the predictions of the hydrodynamic theory are
pretty much insensitive to such small variations in the initial conditions.
(More important is that $\stil(q)$ and $\rho_1(y)$ be consistent with each
other.)

Define an effective susceptibility exponent
\begin{equation}
\gamma_{eff} = -ln\bigl[\stil(0,\epsilon_1)/\stil(0,\epsilon_2)\bigr]/
ln\bigl[\epsilon_1/\epsilon_2\bigr],
\label{eq:gammaeff}
\end{equation}
where the temperatures $\epsilon_1$ and $\epsilon_2$ are close to 
each other in some sense. Then, over this temperature range,
$0.5\leq\epsoverepsf\leq 10$, LBM predicts that $\gamma_{eff}$
varies from 1.39 down to 1.10 for $\alfstar=1$, and 1.47 down to 1.15 for
 $\alfstar=\pi/2$.
Also, for $\alfstar=1$ and $\epsoverepsf=1.5$, and $\alfstar=\pi/2$ and
$\epsoverepsf=2.3$, $\gamma_{eff} = 1.24$, i.e., is exact.

Thus, the LBM theory seems to describe properly the temperature
dependence of critical fluctuations within a window near $\epsoverepsf=1$.
 It is concluded then that using the equilibrium LBM theory, along with 
$\calF$ defined in Sec. \ref{subsec:free_energy} above, to give 
 initial conditions for $\stil(q)$ and $\rho_1(y)$ 
 is acceptable as long as the quenches
are not too deep, $\epsoverepsf > 0.5$.

\begin{figure}
\includegraphics[scale=0.35,trim=0.5in 0.5in 0.3in 0.3in]{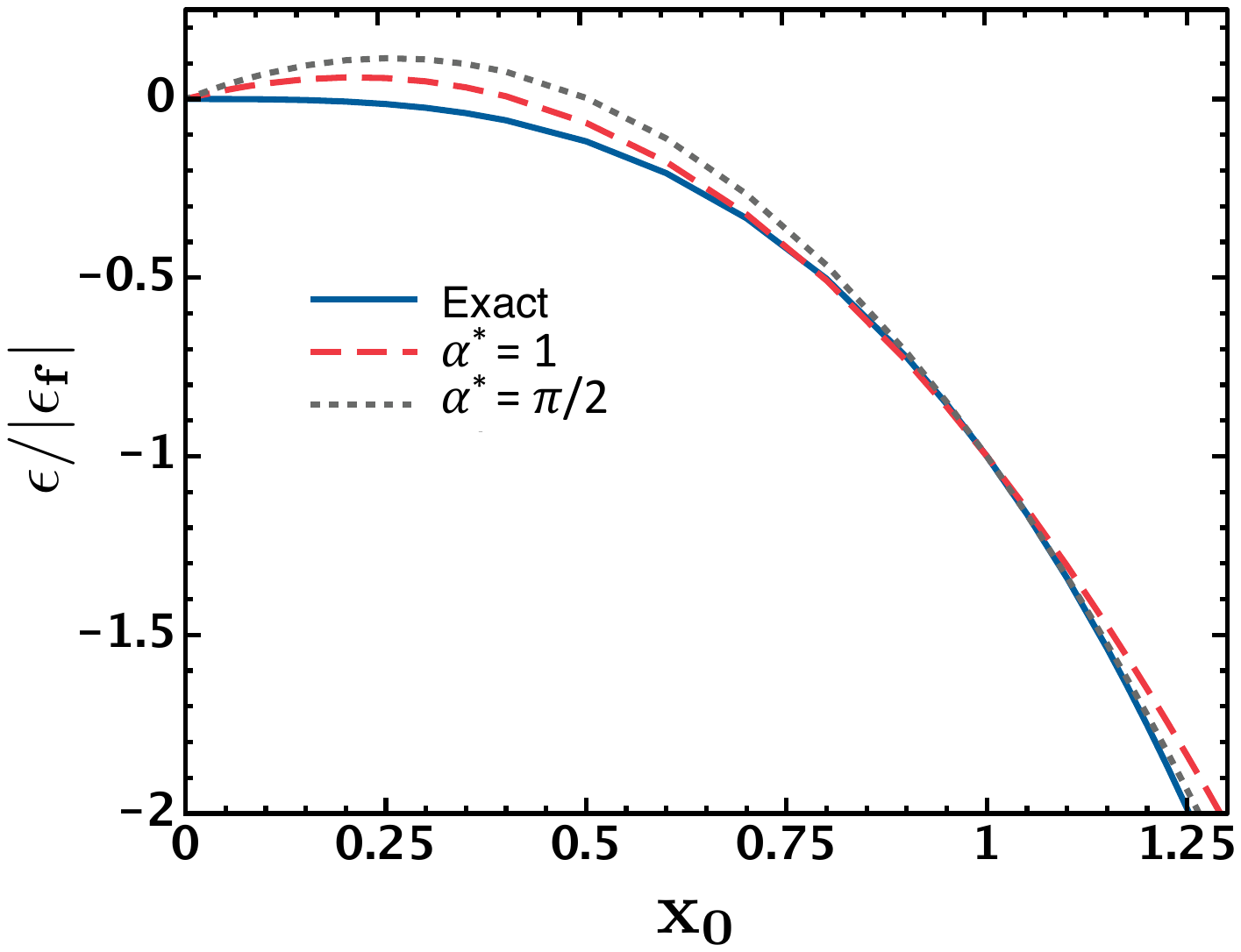}
\caption{\label{figepsvsx0coex}
LBM predictions, using the temperature dependent coarse-grained free energy 
 defined in Sec. \ref{subsec:free_energy}, 
for the coexistence curve $\epsilon(x_0)$
 for two cut-offs $\alfstar$. Also shown is the ``exact''
scaling form in the critical region, $\epsoverepsf = -x_0^{1/\beta}$,
 where $\beta\approx0.33$.
}\end{figure}

Examining the LBM predictions for the two-phase coexistence curve
 $\epsilon(x_0)$ is also illuminating.
Figure \ref{figepsvsx0coex} shows $\epsilon(x_0)$ for positive $x_0$ (the curve
is symmetrical about $x_0=0$) for cutoffs $\alfstar = 1$ and $\pi/2$. Also
shown is the accepted scaling form for a system in the 3-D Ising universality
class: $\epsilon(x_0)\sim -\vert x_0\vert^{1/\beta}$, with $\beta\approx0.33$. 
 Thus, $\epsilon(x_0)$
should have a maximum at $x_0=0$. However, as can be
seen, instead of a maximum at $x_0=0$, the LBM predictions overshoot
 $\epsilon=0$ and have a maximum at around
$x_0=0.2$ and $0.25$ for $\alfstar=1$ and $\pi/2$, respectively.
 Thus, the LBM theory should not be used to describe the initial state for
quenches that are deep, $\epsoverepsf \ll 1$, and off-critical.
Also, contrary to its 
behavior above $T_c$, the LBM predictions for $\gamma_{eff}$ below $T_c$ are 
always below the accepted 3-D Ising scaling value of $1.240$, at best
reaching $0.9$ at $\epsoverepsf=-2.0$.

 On the other hand, the theory's 
predictions for an effective $\beta$ exponent, $\beta_{eff}$, defined
analogously to $\gamma_{eff}$ in Eq. (\ref{eq:gammaeff}) above, are near the
 accepted 3-D Ising value for
temperatures near $\epsilon_f$. For $\alfstar=1$ and $\epsoverepsf=-0.85$,
 and $\alfstar=\pi/2$ and $\epsoverepsf=-1$, $\beta_{eff}$
equals the accepted $\beta$ value of $0.328$.
Also, for temperatures near $\epsilon_f$, the
predictions of LBM for $\epsilon(x_0)$ are in good agreement with the accepted
values in the range $-2.0\leq\epsoverepsf\leq -0.5$. As will be seen below,
the adiabatic theory predicts that the temperature undershoot after an
on-critical quench of 3MP+NE does not go below $-2\epsilon_f$. Thus, if the
quenches are fast, the on-coexistence equilibrium predictions of
the LBM theory should be acceptable.

 It is concluded that if the quenches are not too deep, and are
fast enough so that the fluid spends little time exploring the region near
 $T_c$, then
the temperature dependent coarse-grained free energy defined in 
Sec. \ref{subsec:free_energy} is adequate.

\section{\label{sec:results}Results}
In this section, general predictions of the adiabatic and isothermal
 decomposition theories are discussed, and then the theories are compared
with experiment.
In an unpublished work, Schwartz showed that the original numerical scheme
of KO was not quite right,\cite{schwartz78} leading to erroneous results
for the structure factor at intermediate and late times.
Given this, aspects of the isothermal KO theory by itself will also
be discussed.

\subsection{\label{subsec:genpredictions}General Predictions}
In the isothermal decomposition theory, the equations 
scale completely.  That is, no system or temperature dependent
parameter appears in the theory when the equations are scaled
using parameters appropriate to the critical region.  Thus,
if initial (and quench) conditions are ignored, the theory
predicts that if the experimental data is appropriately
rescaled then the data should superimpose for any binary fluid and any
quench temperature.

However, the adiabatic theory does not scale.  The parameters
appearing in the equation for $\epsilon$, (\ref{depsdteq}), are strongly system
dependent and some, $\alfp$ and $\cp$, are temperature 
dependent. To illustrate the adiabatic results in this section then,
data for 3MP+NE will just be used.  The equation for
$\epsilon$ requires $\alfp,\cp,\rho_c,T_c,dT_c/dP$ and
$\xi_0^\pm$, which, for 3MP+NE, can be obtained from Tables \ref{table3MPNEData}
and \ref{tableExpsAmps}.

The other parameters appearing in the theory are the universal
amplitude $f_1$, the cut-off $\alfstar$, and the cut-off 
dependent free energy parameters $\lambda_i$. As mentioned above,
$f_1$ was set to 0.210, and in Sec. \ref{sec:free_energy} values for
the $\lambda_i$ were computed for various cut-off values. What remains
then is to determine an appropriate cut-off. 

In isothermal decomposition, $\alfstar$ is determined by
requiring that it be large enough so that no unstable modes are
integrated out. In the mean-field theory
of Cahn,\cite{cahn68} the dominant unstable wavevector is at
$q={1/{\sqrt 2}}$,
with the largest unstable mode occurring at $q=1$. Thus,
$\alfstar\geq 1$. While the statistical theory here gives free energy
parameters that are cut-off dependent, this relation roughly holds here too.

 On the other hand, the time dependent
inverse susceptibility, $A(t)$, appearing in the equation of motion
for the structure factor, Eq.(\ref{dshatdteq}) does not vary
with the wavevector $k$ of the mode. In other words, the LBM ansatz
produces a mean-field form for this time dependent inverse 
susceptibility.\cite{langer75} The goal then should be to include
as few concentration modes as possible into this mean-field approximation,
that is, to make $\alfstar$ as small as possible. A good compromise between
these opposing needs is to follow LBM and just let $\alfstar=1$ for
 isothermal decomposition.

In adiabatic decomposition, the wavevector of the largest unstable mode 
will depend upon the degree of temperature undershoot. What has been done
here is set $\alfstar$ and then examine the temperature undershoot at a short
time after the end of the quench, say, $\tau=0.1$. The inverse of the
equilibrium correlation length for that temperature at that time was then
identified to be the minimum cut-off value, in analogy with the isothermal case. For the
quenches considered here, it was found that setting $\alfstar=1.4$ 
was reasonable. For consistency, this cut-off was also used
for the isothermal runs.

With the parameters in the equations determined, a quench 
is specified by the initial reduced temperature
 $\epsilon_i$, the pressure change $\Delta P$, and the
quench time $\tau_{quench}$.  The final temperature was
 determined by integrating Eq.\ (\ref{32j}). The quench time varied
with experiment, but was either known or could be deduced.
 
\begin{figure}
\includegraphics[scale=0.31,trim=0.2in 0.4in 0.3in 0.2in]{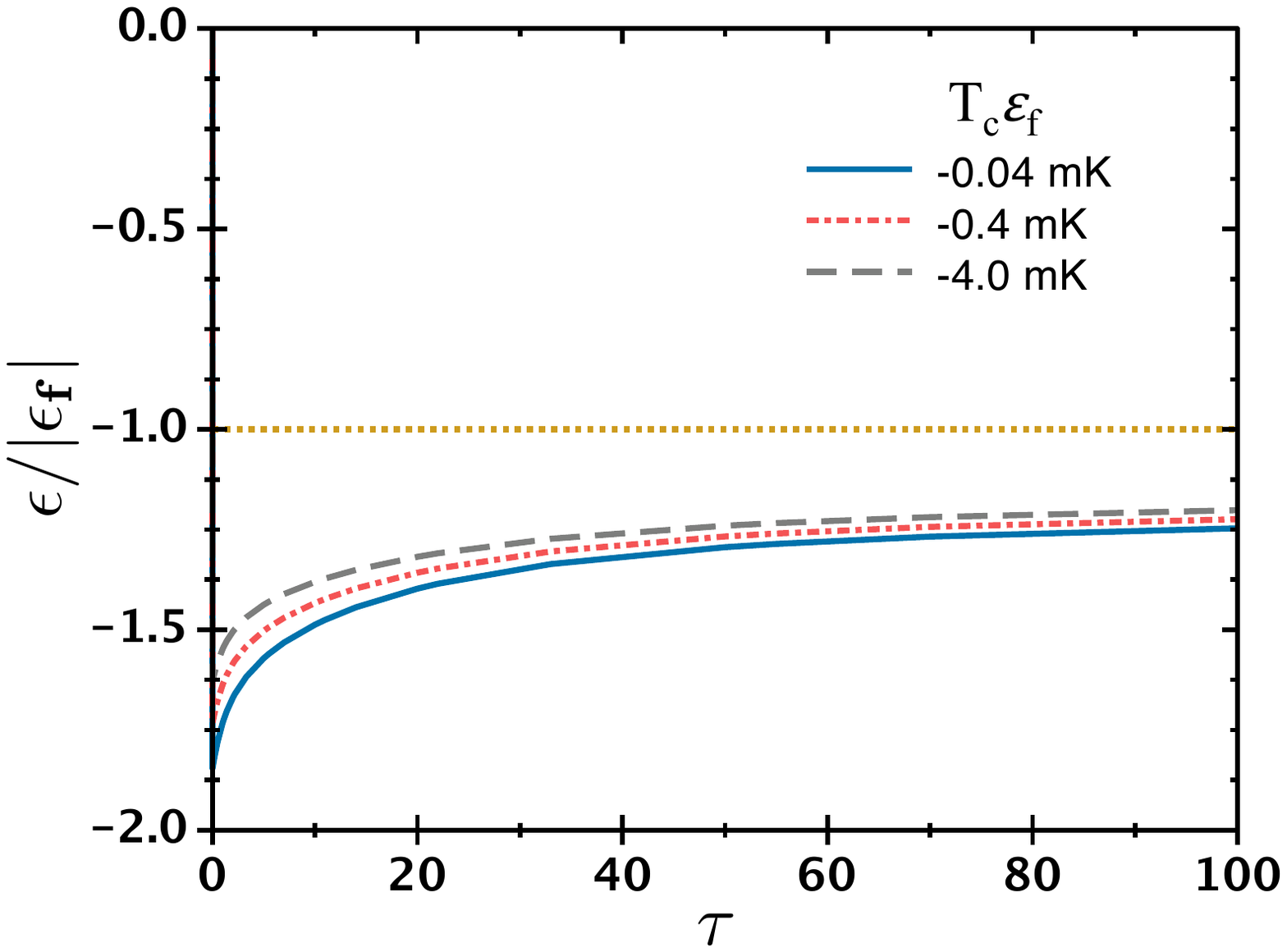}
\caption{\label{figepsvstau}
Scaled temperature $\epsilon/|\epsilon_f|$ as a
function of scaled time $\tau$ for three adiabatic runs.
The final temperatures $T_c\epsilon_f$ are indicated in the figure and the
initial temperatures are $\epsilon_i= -10\epsilon_f$. The straight line 
denotes the final equilibrium temperature.
}
\end{figure}

Figure \ref{figepsvstau} shows the scaled temperature,
 $\epsoverepsf$ as a function of the scaled time $\tau$ for
three adiabatic runs ending at the temperatures 
$T_c\epsilon_f= -0.04$ mK, $-0.4$ mK and $-4.0$ mK, with
initial temperatures $\epsilon_i = -10\epsilon_f$.
The quenches were on-critical so $x_0=0$.
The scaled quench time $\tau_{quench}$ has been set to be
$0.01$; thus the initial temperature drop does not appear on
the graph. Clearly, the temperature undershoot
is large; the temperature reached immediately after the
quench is roughly $-1.8 |\epsilon_f|$, with the smaller
final temperatures giving the greater undershoot. As can be seen, there is a
sharp rise from this minimum at early times $\tau < 5$, and then a gradual rise later. 
This qualitative behavior has been seen by Milchev et al.\ in a 2-D Ising simulation of adiabatic decomposition.\cite{milchev94}  Note
also that there is not much difference in the scaled
temperature trajectories even though the final scaled
temperatures differ by a factor of 100.

\begin{figure}
\includegraphics[scale=0.58,trim=0.2in 0.4in 0.3in 0.3in]{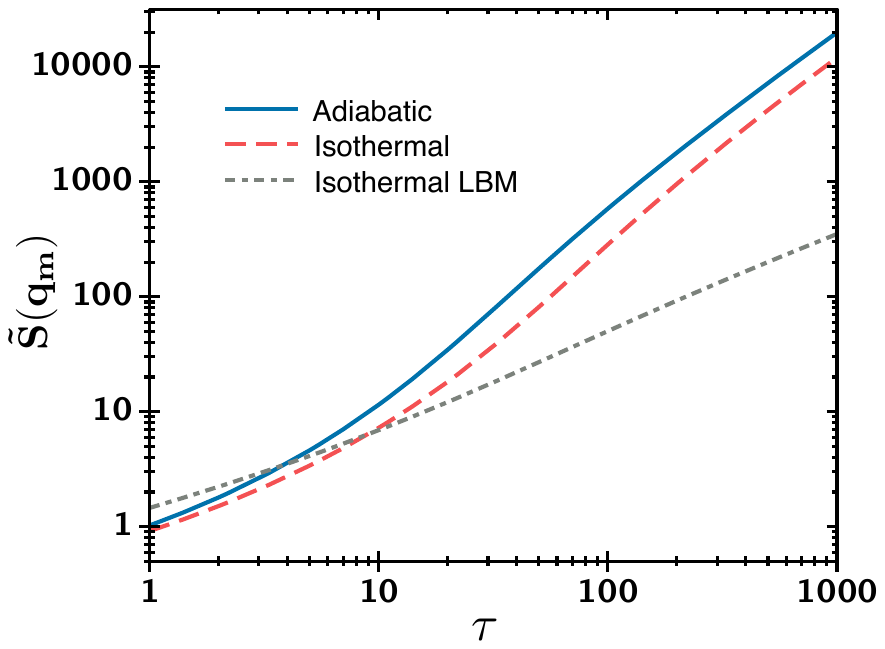}
\caption{\label{figsqmvstau}Scaled structure factor peak $\stil(q_m)$
as a function of scaled time $\tau$ for an adiabatic and isothermal run
of the KO theory. Also shown are predictions of an isothermal run from an
 LBM version of the theory.
 The adiabatic run is the same as the middle one in Fig.\ref{figepsvstau}:    $T_c\epsilon_f = -0.4$ mK and $\epsilon_i/\epsilon_f=-10$; the isothermal 
runs have the same ratio of initial to final temperature. Results for
adiabatic runs for other temperatures shown in Fig.\ref{figepsvstau} gave
peak values that differed at most by $5\%$ from the $T_c\epsilon_f=-0.4$ mK run
here. }
\end{figure}

\begin{figure}
\includegraphics[scale=0.32,trim=0.5in 0.5in 0.3in 0.3in]{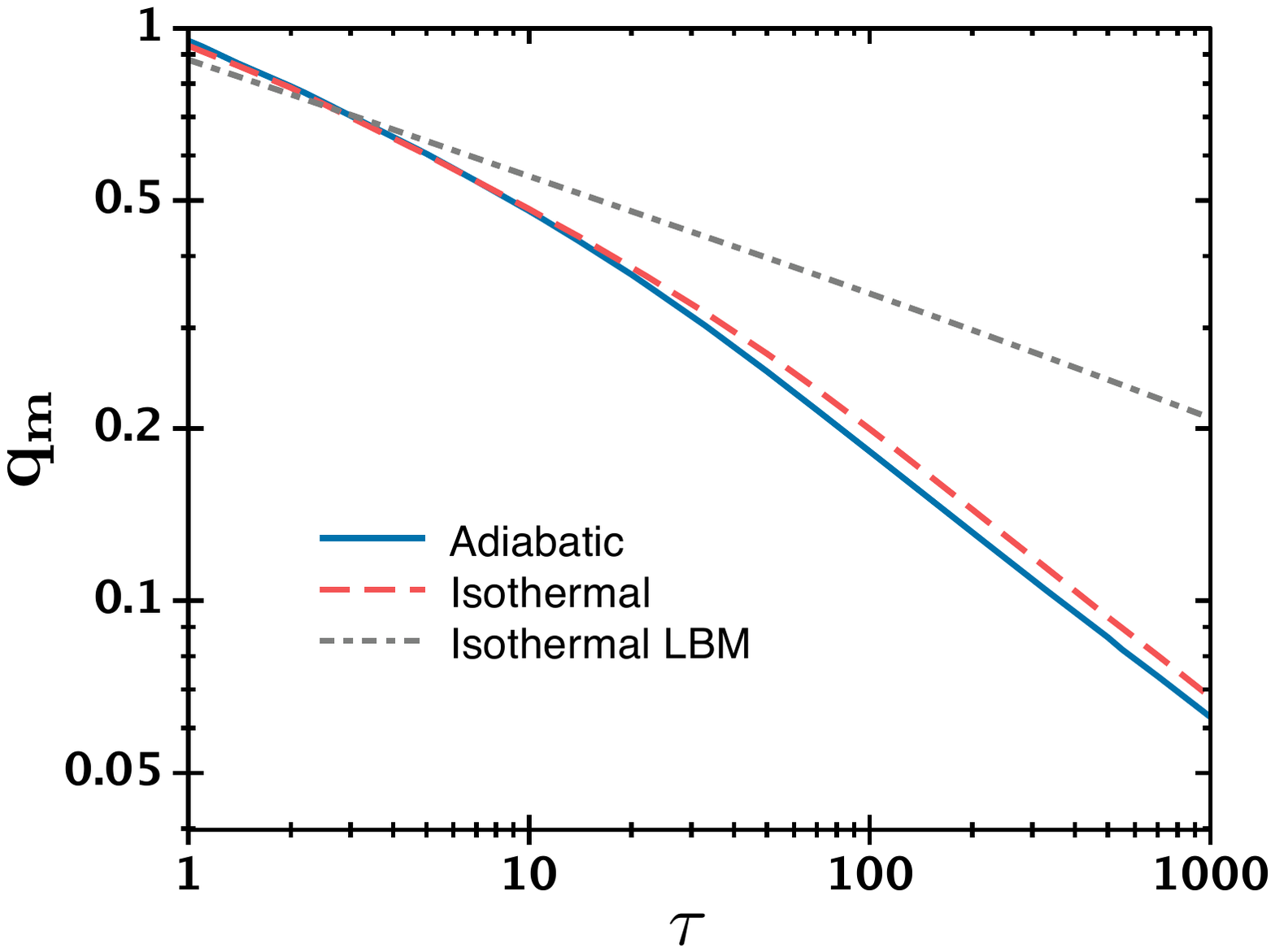}
\caption{\label{figqmvstau}
Scaled peak wavevector $q_m$ as a function of scaled time $\tau$ 
for an adiabatic and isothermal run of the KO theory. Also shown are 
predictions of an isothermal run from an LBM version of the theory.
The conditions are the same as for the results shown in Fig.\ref{figsqmvstau}.
Results for the other adiabatic runs in Fig.\ref{figepsvstau} were essentially
identical to the one shown here.
}\end{figure}

Figures \ref{figsqmvstau} and \ref{figqmvstau} show results for the scaled 
peak intensity, $\stil(q_m)$, and scaled peak wavevector,
 $q_m$ as functions of the scaled time $\tau$. Results of the middle adiabatic
quench in Figure \ref{figepsvstau}, $T_c\epsilon_f=-0.4$ mK, are shown along
with results from an isothermal run with the
same ratio of $\epsilon_i/\epsilon_f = -10$. Also shown are results from an
``LBM'' version of the theory in which $\sigma(q)$, Eq.(\ref{sigmaqeq}), is
set to 1 and the second term in Eq. (\ref{dstilqdteq}) due to the hydrodynamic
operator $\calL_2$ is dropped. Setting $\sigma(q)=1$ assumes all modes
have equilibrated, which clearly is not the case, so that value should
be considered an upper bound.  In Figure \ref{figsqmvstau} it
 can be seen that 
the temperature undershoot causes $\stil(q_m)$ for the
adiabatic quench to grow initially more rapidly than
the isothermal quench.

Interestingly, $\stil(q_m)$ for the adiabatic quench in Figure \ref{figsqmvstau}
never differed from the peak height of the other two adiabatic quenches in
Figure \ref{figepsvstau}
(not shown) by more than about $5\%$ at 
late times even though there is a spread of two orders of
magnitude in their final temperatures.  This weak violation of
scaling is caused by the weak divergence ($\alpha=0.105$) 
of $\alfp$ and $\cp$. On the other hand, $\stil(q_m)$ for the
isothermal quench differs by at least a factor of 2 from the adiabatic runs.
At very early times, the LBM prediction is greater than either version
of the KO theory, due presumably to the overestimation of the transport
function $\sigma(q)$. At later times, LBM lags appreciably
behind KO as expected.

In Figure \ref{figqmvstau} it can be seen that initially the adiabatic run
predicts a larger peak wavevector than the isothermal one, but the 
adiabatic quickly overtakes the isothermal. This is behavior has also been seen by Milchev et al.\cite{milchev94} It is due to the adiabatic run quenching to
a lower temperature causing the fluid to coarsen on a smaller lengthscale and at
a faster rate (as stated before, the characteristic decomposition time $\sim\vert\epsilon\vert^{-1.94}$ in liquids). At large times, the time dependence of both types of runs are the same, though the adiabatic one gives gives a slightly smaller $q_m$.
Thus, one may not be able to determine any great discrepancy between the isothermal theory and experiment if one looks only at the peak wavevector.

While KO, like LBM, was created to describe the early stage of decomposition,
it is interesting to examine its behavior at later times, if only to estimate when it breaks down. 
 Define time dependent exponents,
$a_q$ and $a_s$, so that $q_m\sim \tau^{-a_q}$ and 
 $S(q_m)\sim \tau^{a_s}$ at any $\tau$. 
 For $q_m$, $a_q$ increases monotonically at early
times, but appears to approach a constant for $\tau > 100$. It was found that
in the isothermal case KO and LBM predict that $a_q \approx 0.47$ and 0.22,
 respectively, for
$100<\tau\leq 1000$.  At larger times, $10^3< \tau < 10^4$, the KO
value decreases slightly to $0.46$. The exponent changed only slightly with
cut-off, $a_q\approx 0.47$ and 0.46, for $\alfstar=1$ and
$\alfstar=\pi/2$, respectively,
 for the largest times examined, $10^3< \tau\leq 10^4$.
Note that at $\tau\approx 10^4$, $q_m\approx 0.02$.
In absolute terms, $q_m$ varied less than 5\% with cut-off out to $\tau=100$.
The adiabatic theory predicted about the same behavior for $q_m$ at these late times.

As mentioned above, the KO theory applies to fluid flow at low Reynolds number, 
that is, when the viscous term in the Navier-Stokes equation is much larger than the inertial one. In this limit, it is expected that the
dominant mechanism in the very late stages of coarsening
yields $a_q = 1$.\cite{siggia79,sain05} As a consequence,
the KO value for $a_q$ can then at most
be considered valid for an intermediate stage of phase separation. Given the results shown in Figure \ref{figsqmvstau}, this limit is $\tau\approx 100$.

The effective time exponent of $\stil(q_m)$, $a_s$, also increases at early
times; however at $\tau\approx 100$ for KO it reached a maximum of 1.8 and
then dropped slowly, reaching a value of 1.46 at the largest times examined,
$\tau \approx 10^4$. On the other hand,
experiments have shown that $a_s$ increases monotonically with time,
eventually approaching a constant.\cite{chou79,wong78}
So this slowing in the growth of $\stil(q_m)$ seems to indicate a gradual
breaking down of the theory. The peak height was more sensitive to the
cut-off with the maximum of $a_s$ for KO being 2.1 and 1.7 for $\alfstar=1$
and $\pi/2$, respectively. On the other hand, this maximum for KO always 
occurred when $q_m\approx 0.2$. (For LBM it occurs for $q_m\approx 0.35$.)
 This value of $q_m$ corresponds
to an average fluctuation size of $\pi\xi_f/q_m\approx 16\xi_f$. In 
Cahn-Hilliard theory \cite{cahn58}, the equilibrium interface separating
two phases has a width of around $4\xi_f$. Thus, at this time sharp 
interfaces will be forming, which the LBM and thus KO theories cannot
describe.\cite{langer75}
At $\tau = 100$, $\stil(q_m)$ for the isothermal KO theory was 458, 285 and 239
for $\alfstar = 1$, 1.4 and $\pi/2$, respectively, so the cut-off dependence of
the theory seems to decrease as the cut-off is increased.

Scaling theory\cite{binder74,furukawa85} predicts at late times that the
function $F(x) = q_m^3\stil(q = x q_m)$ becomes constant. Interestingly,
the peak of this function is almost a constant within the KO theory:
$F(1)\sim\tau^{\zeta_F}$, with $\zeta_F \approx 0.07$ at late times. 
This trend of the theory persists at least out to $\tau\approx 10^4$.

\subsection{\label{subsec:compareexpt}Comparison With Experiment}
In this section the adiabatic and isothermal theories will be
compared with light scattering data of BC\cite{bailey93}.
\begin{figure}
\includegraphics[scale=0.66,trim=0.2in 0.3in 0.3in 0.4in]{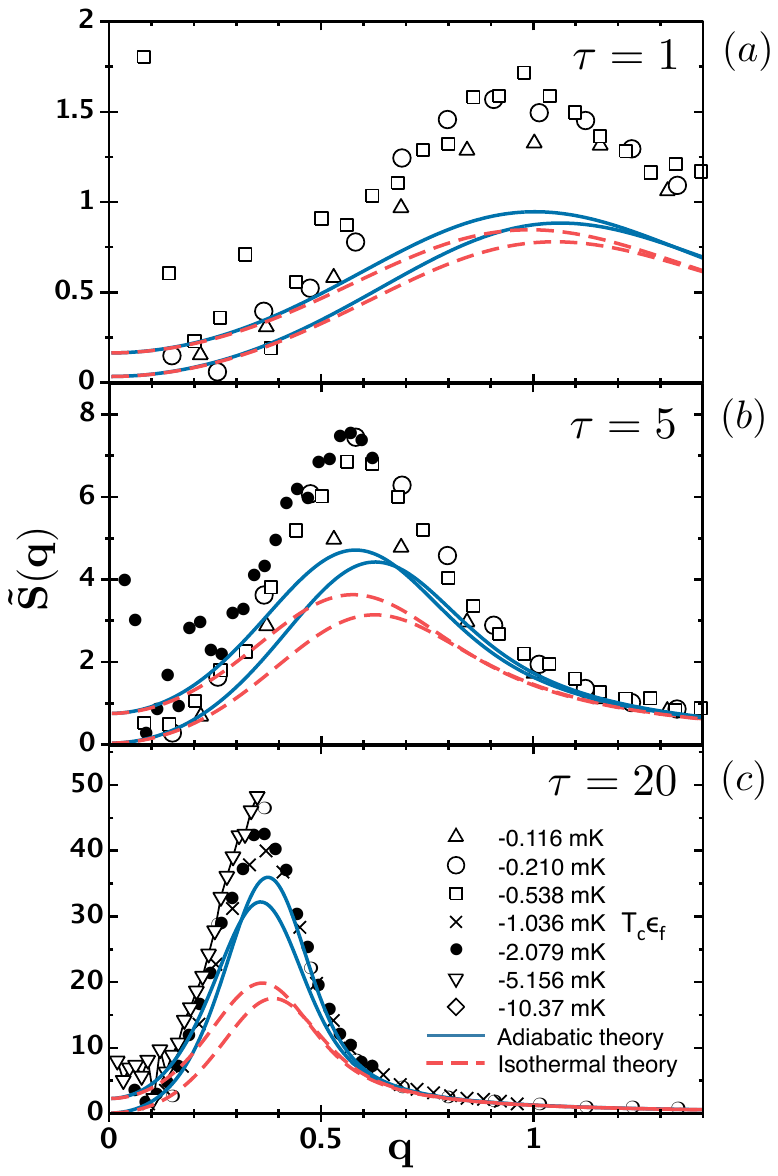}
\caption{\label{figsqvsq1n5n20}
(a) Scaled structure factor $\stil(q)$ as a function
of scaled wavevector $q$ at scaled time $\tau=1$; (b) the same for $\tau=5$;
 and (c) the same for $\tau=20$.
The initial temperature for all
quenches was $T_c\epsilon_i$= 10 mK. The symbols denote the
experimental results of Bailey and Cannell as follows:
$T_c\epsilon_f$ = $(\bigtriangleup)$ -0.116 mK,
 $(\bigcirc)$ -0.210 mK, $(\Box)$ -0.538 mK, $(\times)$
-1.036 mK, $(\bullet)$ -2.079 mK,
$(\bigtriangledown)$ -5.156 mK, and $(\diamond)$ -10.37 mK.
The blue solid and red dashed curves denote results of the adiabatic
and isothermal theories, respectively. For the isothermal runs for each time,
the upper curve denotes results for the deepest quench shown in
that time frame, while the lower one is for the shallowest. This meaning
also holds for the adiabatic runs in (a) and (b), but the reverse
is true for (c).
}
\end{figure}
\begin{figure}
\includegraphics[scale=0.51,trim=0.1in 0.3in 0.3in 0.4in]{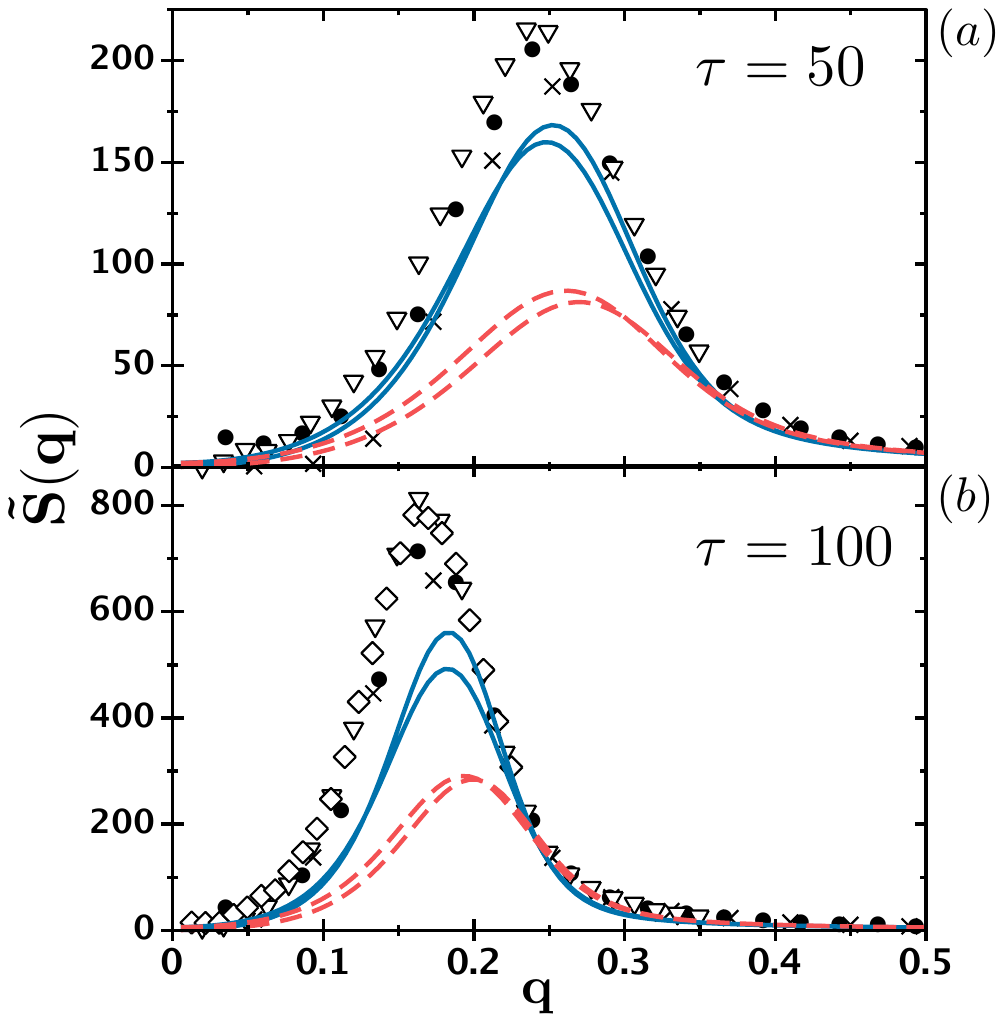}
\caption{\label{figsqvsq50n100}
(a) Scaled structure factor $\stil(q)$ as a function
of scaled wavevector $q$ at scaled time $\tau=50$; and (b) the same for $\tau=100$.
The meaning of the symbols, curves and conditions are the
same as in Fig.\ref{figsqvsq1n5n20}(c).
}
\end{figure}

As stated above, the experimental quenches of BC were for on-critical mixtures so $x_0=0$. To compare the adiabatic theory with these quenches, the
thermodynamic quantities $T_c,dT_c/dP,\rho_c,\alfp$ and $\cp$ are needed. All can be found using Tables \ref{table3MPNEData} and \ref{tableExpsAmps}. Also needed are the two-phase values for $\xi_0$ and $\eta_s$. The former can be deduced from data in the same tables to be $\xi_0^-=1.13\AA$. 

As mentioned above, the hydrodynamic shear viscosity, $\eta_s$, is not constant but is a singular function of $\epsilon$.  In addition,  the two-phase value of 
$\eta_s$ has not been determined, and at present there is no definite relation 
between the one and two-phase amplitudes.  However, since
the scaling form for $\eta_s$ is so weakly singular and 
the quenches are expected to be fast, 
$\eta_s$ was simply set equal to its one-phase value at $\xi=\xi_f$, 
i.e., $\eta_s\simeq{\bar\eta}(Q_0\xi_f)^{z_\eta}$,
where $\bar\eta$, $Q_0$ and $z_\eta$ are given in Table \ref{table3MPNEData}.

For each quench, the initial temperature, $\epsilon_i$, and the pressure change,
 $\Delta P$, are known. In the constant entropy approximation of the theory here, the final temperature, $\epsilon_f$, was determined by just integrating
Eq.\ (\ref{32j}). 
Re-evaluation of the critical properties of 3MP+NE by
BC allows us to ignore any uncertainty in $\epsilon_f$.
The experimental intensity data was scaled by BC\cite{baileyThesis}.

In Figures \ref{figsqvsq1n5n20} and \ref{figsqvsq50n100},
the scaled structure factor $\stil(q)$ is shown as a function of the scaled
wavevector $q$ for various scaled times $\tau$.
The quenches shown all begin at $T_c\epsilon_i\approx 10$ mK and
have final temperatures that range from $T_c\epsilon_f= -
0.116$ mK to $-10.37$ mK. The solid and dashed curves
denote results for the adiabatic and isothermal theories, respectively,
while the points represent data of BC. For the isothermal runs for each time, the uppermost curve
denotes the result for the deepest quench shown and the lower one
denotes the shallowest. This meaning also holds for the adiabatics runs
for $\tau=1$ and $5$, shown in Figures \ref{figsqvsq1n5n20}(a) and \ref{figsqvsq1n5n20}(b), but the reverse becomes true for larger times. The effect
of a finite quench time is included in these results; the scaled quench time
$\tau_{quench}$ ranged from $2\times 10^{-3}$ for the
shallowest quench to $11$ for the deepest.

At a very early time, $\tau=1$, the prediction of the adiabatic theory
for the peak height $\stil(q_m)$ lags behind the data by a factor of 2.
However, at later times, $\tau=20$ and $50$, the agreement with
experiment is very good, within $20\%$. At the largest time for which
data is available, $\tau=100$, the adiabatic theory
appears to start lagging behind the data again, with the
difference being $30\%$. On the other hand, the predictions of the adiabatic 
theory for the peak wavevector $q_m$ are within a few percent of the
data at all times.

The LBM and thus KO theories are expected to work best at early times. For
example, Mainville et al.\ obtained good agreement throughout the early stage,
albeit with some fitting, between their experimental scattering data and the
LBM theory.\cite{mainville97}
 Therefore, the disagreement between the adiabatic theory 
and experiment for the peak height $\stil(q_m)$ at very early times is
perplexing. One possibility is that setting the cut-off to a finite value
removes the relaxation of high wavevector modes, $q>\alfstar$, right after the
quench. The relaxation of these modes then couple to lower $q$ ones, increasing
their relaxation, like a wave moving through $q$-space. To examine this
hypothesis, the cut-off of the theory was varied from $\alfstar =1$ to
$\alfstar = \pi/2$. It was found that $\stil(q_m)$
 at $\tau=1$ varied by only 9\% and 3\% for the adiabatic and isothermal
theories, respectively, for this cut-off range.

Another possible explanation is that the adiabatic theory has 
underestimated the temperature undershoot. Though whatever the cause, further research is needed.

The disagreement between theory and experiment at late times, is that at
$\tau=50$, $q_m\simeq 0.2$, which, as has been discussed above, appears to
be where the KO theory begins to break down.
The isothermal theory predicts a peak height and peak wavevector that
lags behind experiment at all times, the difference in $\stil(q_m)$ becoming
over a factor of three at the latest times.

Note that the theory results shown in Figures \ref{figsqvsq1n5n20} and
 \ref{figsqvsq50n100} are not quite the same as those in a previous
report of the adiabatic theory, Ref.\cite{donley93}.
 One reason is that in Ref.\cite{donley93} the ``double gaussian" approximation was used to solve for the time evolution of $\rho_1(y)$.
As mentioned above, while this approximation is more computationally efficient than solving the full PDE for $\rho_1(y)$, Eq.(\ref{drho1ydteq}), it tends to overestimate the growth of $\stil(q)$ for times such that $q_m\geq 0.3$. Advances in computer power in the years since 
Ref.\cite{donley93} was published have made this approximation unnecessary.
A second reason is that the form for the scaled free energy density
$\phi(x)$ was different in the previous work. This previous form was constructed
to satisfy a constraint in the limit of the cut-off $\alfstar\rightarrow 0$
 (see Ref.\cite{donleyThesis} for a detailed description). It was subsequently concluded that the added complexity to $\phi(x)$ needed for this constraint outweighed any improved accuracy of the theory, and so here the standard
``$\varphi^4$" form for $\phi(x)$ was used instead.

As mentioned above, the isothermal theory predicts that if the
fluid is in the critical region, the experimental data is scaled
properly, and the scaled initial conditions and quench times are the
same, then the scaled time evolution of any experimental run should be
identical. 
It is interesting then whether the experimental data of BC
show any violation of this scaling. Consider two experimental runs of
BC with final temperatures $T_c\epsilon_f=
-2.079$ mK and -0.202 mK.\cite{baileyThesis}  The initial temperatures
were both at $\epsilon_i \simeq 5\epsfabs$ to eliminate the
effect of initial conditions.\footnote{The
times for these quenches were estimated from other data by 
assuming that the rate of change of the pressure was the same for
all quenches.}
At $\tau=10$, $\stil(q_m)$ was measured to be 16.6 and 18.8 for
the first (-2.079 mK) and second (-0.202 mK) quench, respectively. The
 adiabatic theory predicts that
$\stil(q_m)=11.4$ and 12.3, for the first and second quench, respectively,
while the isothermal theory predicts that 
$\stil(q_m)=7.6$ and 7.7 for those quenches.
Both the adiabatic and isothermal theories predict that $q_m\approx 0.47$
in agreement with both experimental runs.
While the experimental violation of scaling is not large,
the difference in $\stil(q_m)$ for the two runs being 12\%, the
trend is in agreement with the adiabatic theory, which predicts
a difference of 8\%. So though there is certainly scatter in the data, this
agreement at the least is suggestive that
the temperature change during decomposition is appreciable for
this fluid.

\section{\label{sec:summary}Summary and Discussion}
In summary, the KO-LBM theory of spinodal decomposition
in binary fluids was generalized to model experimental scenarios in which
the fluid is quenched by changing the pressure and the subsequent phase
separation occurs adiabatically.

The central idea of the approach here is that the coarse-grained free 
energy, $\calF$, which governs the time evolution of the slowest
modes, is constructed in a manner that creates a natural split in
the degrees of freedom of the system. Those fast degrees of freedom that have
been integrated out contribute to $\calF$, but also to a free energy,
$F_r$, that is independent of the configuration of the slow modes.
It was shown that the fast degrees of freedom, through $F_r$, are able to
act as a thermal reservoir for the slow modes.
Any global constraint though, such as constant energy or entropy, indirectly
relates the state of the reservoir, and thus its temperature, to the particular
 state of the slow modes. However, it was argued
 that these states need be related only in an average sense. With that approximation,
an equation of motion for the average reservoir temperature was derived,
it playing the same role as the assumed constant global system
temperature in previous isothermal theories of decomposition.

In other words, if in a system there exists degrees of freedom that are able to sample their accessible states on timescales smaller than the characteristic change of the slower degrees of freedom, then these fast degrees of freedom can act as a reservoir to the slow ones. This model of nonequilibrium processes has become a common one.\cite{kurchan05,mosgaard13,bouchbinder13}

The extension of the isothermal theories of KO and LBM to adiabatic conditions
then consisted of: this equation of motion for the reservoir
temperature; estimates for various reservoir thermodynamic derivatives,
such as the heat capacity, which appear in the temperature equation; and 
a specification of a temperature dependent coarse-grained free energy.

 This ``adiabatic'' theory was then applied to
an on-critical mixture of 3MP+NE.  It was shown that the
temperature change during decomposition is appreciable and 
accelerates the coarsening.
The adiabatic and previous isothermal theories were then compared
quantitatively, with no adjustable parameters, with data of Bailey and Cannell
 on 3MP+NE for the structure 
factor at various times during the early stage of decomposition.  It was shown
 that there is a definite lack of agreement between the data and previous
theory for the structure factor peak height, and that the adiabatic theory
accounts for a substantial amount of this difference.
The adiabatic theory also improves the agreement with experiment for the
wavevector, $q_m$, of the structure factor peak.
Differences between theory and experiment though indicate that the
adiabatic theory may still be underestimating
the effects of temperature changes during decomposition for 3MP+NE.
Further research is needed to determine the cause.

The large temperature change during decomposition predicted for 3MP+NE is due 
partly to the size of the singular term in the isobaric heat capacity
compared to the background term
(see Table \ref{table3MPNEData}). Another binary fluid, isobutyric acid and
water, has a much smaller singular term, and for it, at the same reduced
temperatures as the Bailey and Cannell experiments, the predictions of the
adiabatic and isothermal decomposition theories are essentially
the same.\cite{donleyThesis}
It is possible though that there are other binary fluids with even
larger relative singular contributions to their heat capacity than 3MP+NE,
making the adiabatic effect even more pronounced.

The adiabatic theory could possibly be extended to temperatures
outside the critical region, e.g., for mixtures with longer range 
interactions such as polymers. However, if 3-D Ising critical scaling no longer
holds, the isothermal KO/LBM theory itself becomes temperature dependent through
at least the parameter $f_1$ (see Sec. \ref{subsec:free_energy}). So it is 
unclear how the predictions of this extended adiabatic theory would differ from
the near-critical one developed here, especially as
thermodynamic quantities such as the heat capacity are system dependent.

The behavior of the isothermal KO theory at later times was also
 analyzed. It was found for times $10^2 < \tau \leq 10^4$, that $q_m$ scaled
as $\tau^{-a_q}$, with $a_q\approx0.46$.

While off-critical quenches were not examined here,
it is expected that the adiabatic effect to be less for them since the heat
released during phase separation should be largest for an on-critical mixture,
it being roughly proportional to $1-x_0^2$, using Eq. (\ref{eq:Ecgavg}).

Interestingly, it was shown in Section \ref{subsubsec:adiabatic_system} that
 the entropy increase during this adiabatic decomposition is
well approximated as zero.
That is, in the model here, the temperature rise from phase
separation exactly compensates for the temperature undershoot caused by the
incomplete relaxation of the slow modes during the quench, so that the final
temperature reached is as if the whole process had been reversible. In that
manner, if a fluid were quenched, allowed to phase separate at least
partially, and then the pressure were reversed, the fluid upon re-mixing should
reach a temperature very near its initial value. A similar two-step experiment
was done by Siebert and Knobler in their study of nucleation.\cite{onuki02,siebert84}
While the arguments leading to this prediction of a (almost) constant entropy
decomposition relied partly on the system being a near-critical binary fluid,
it might be more general. Answers are left to future research.

\begin{acknowledgments}
I thank James Langer for many helpful discussions,
and Arthur Bailey and David Cannell for suggesting this problem and
many discussions. I also thank John McCoy for some discussions of thermodynamic fundamentals, and Craig Pryor and Jonathan Simon for peripheral conversations.
\end{acknowledgments}

\appendix*
\section{\label{sec:app}Coarse-grained energy}
In this appendix it is shown that in the critical region, the dominant contribution to the average coarse-grained energy, $\calEavg$, comes from the heat of mixing.
First, except when stated otherwise, all quantities are evaluated on-coexistence at the final equilibrium temperature, $T_f$, with a reduced cut-off $\alpha^* = 1$.
Now, using results of section \ref{subsec:free_energy}, $\calEavg$ is the sum of
three terms: the gradient energy,
\begin{eqnarray}
\langle\calE_K\rangle=&&{KV\over 2}\Bigl\langle (\nabla c(\bfr))^2\Bigr\rangle\\
=&& {KV\over 4\pi^2} \int_0^\Lambda dk k^4 {\hat S}(k) \nonumber \\ 
 \simeq&& 0.003\biggl({k_BT_f V\over \xi_f^3}\biggr),\nonumber
\label{eq:Ecgkapavg}
\end{eqnarray}
the heat of mixing,
\begin{eqnarray}
\langle\calE_{mix}\rangle =&& -V\Bigl\langle T^2{\partial (f(c)/T)\over \partial T}\Bigr\rangle \\
=&& -\bigl[(\lambda_2-\lambda_0){1\over \epsfabs} + \lambda_0\bigr]f_1\langle x^2\rangle { k_B T_f V\over \xi_f^3} \nonumber \\ 
\simeq&& -{0.08\over\epsfabs}\biggl({k_BT_f V\over \xi_f^3}\biggr),\nonumber
\label{eq:Ecgmixavg}
\end{eqnarray}
and the fluid kinetic energy,
\begin{eqnarray}
\langle\calE_\bfv\rangle =&&\ {1\over 2}\rho V \langle \bfv(\bfr)^2\rangle \\
\approx&& \ {1\over2}\rho KV\int^\Lambda {d\bfk_1\over (2\pi)^3}{1\over \eta_s k_1^2}\int^\Lambda {d\bfk_2\over (2\pi)^3} \bfk_2\cdot
{\bf\hat T}(\bfk_1)\cdot{\bfk_2} \nonumber \\ 
\times&& \bigl[k_2^2 - (\bfk_1 - \bfk_2)^2\bigr]
\bigl[K k_2^2 + A(t)\bigr] {\hat S}(\vert\bfk_1-\bfk_2\vert){\hat S}(k_2) \nonumber \\
=&&\ \biggl({k_BT_f V\over \xi_f^3}\biggr)\Omega I_{\bfv} \nonumber
\label{eq:Ecgvavg}
\end{eqnarray}
Here, $I_{\bfv}$ is a scaled integral and $\Omega \equiv {\rho k_BT/(\eta_s^2\xi)}$ is a dimensionless number.
To evaluate Eq.(\ref{eq:Ecgvavg}), the same overdamped approximation for $\bfv$ done in the KO theory was used, except that the noise term was dropped as it gives a term that is constant in the critical region. So, $\bfv\approx {\bf T}\cdot (\mu\nabla c)$.
Similarly, four-point averages arising in Eq.(\ref{eq:Ecgvavg})
were evaluated in the manner done by KO for the equation of motion
 of ${\hat S}(k,t)$, Eq.(\ref{dshatdteq}).

The quenches for BC had final temperatures, $\epsfabs\sim 10^{-5}$. Then, $\Omega\approx 10^{-5}$. The scaled integral $I_{\bfv}$ was zero in equilibrium. Out of equilibrium during a typical quench it was non-zero, but remained very small, $\sim 10^{-6}$.
So, $\langle\calE_K\rangle/\langle\calE_{mix}\rangle \sim 10^{-6}$ and $\langle\calE_\bfv\rangle/\langle\calE_{mix}\rangle \leq 10^{-15}$, and so only $\langle\calE_{mix}\rangle$ need be considered. It is interesting that while advection dominates the kinetics of unmixing, in the critical region its contribution to the energetics is negligible.

\bibliography{paper}
\end{document}